\documentclass{ar-1col}

\usepackage{url}
\usepackage[numbers,sort&compress]{natbib}

\usepackage{xspace}
\usepackage{graphicx}
\usepackage{amssymb}
\usepackage{enumitem}
\usepackage[svgnames]{xcolor}
\usepackage[bookmarks=false,colorlinks=true,linkcolor=DarkBlue,citecolor=DarkBlue]{hyperref}
\usepackage{subfig}
\usepackage{rotating}
\usepackage{units}
\usepackage{amsmath}
\usepackage{bm}     % nice math bold italics
\usepackage{lineno}
\usepackage{listings}
\usepackage{lmodern} % suppress font warnings
\usepackage[normalem]{ulem}
\usepackage[section]{placeins}
\usepackage{hepunits}
\usepackage{hepparticles}
\usepackage{cancel}
\usepackage{hepnames}
\usepackage{mathtools}
\usepackage[capitalise]{cleveref}
\usepackage{braket}
\usepackage{slashed}
\usepackage{multirow}

\usepackage[Export]{adjustbox}

\def\g{\gamma}
\def\t{\tau}

\def\D{\Delta}
\def\G{\Gamma}
\def\O{\Omega}

\makeatletter
\newcommand*\dotp{\mathpalette\bigcdot@{.5}}
\newcommand*\bigcdot@[2]{\mathbin{\vcenter{\hbox{\scalebox{#2}{$\m@th#1\bullet$}}}}}
\makeatother

\newcommand{\tctza}{\ensuremath{(t_c-t_0)}}
\newcommand{\tctzb}{\ensuremath{\Big(\frac{-t_0}{t_c-t_0}\Big)}}

\def\Dslash{D\hskip-0.65em /}
\def\Dslashe{D\hskip-0.5em /}
\def\tsep{t_{\rm sep}}
\def\tmin{t_{\rm min}}

\iftrue
\usepackage{feynmp-auto}
\newcommand{\Umunu}{
\hspace{-5em}
\scalebox{0.93}{
\raisebox{-.3em}{
\begin{minipage}{0.14\textwidth}
\begin{fmffile}{plaquette}
\fmfset{arrow_len}{2.5mm}
\fmfframe(20,20)(100,10){
\begin{fmfgraph*}(50,40)
 \fmfleft{l0,l1}
 \fmfright{r0,r1}
 \fmf{fermion,l.d=0.1,l.s=right, label=$U^\dagger_\nu(x)$}{l1,l0}
 \fmf{fermion,        l.s=bottom,label=$U_\mu(x)$}{l0,r0}
 \fmf{fermion,        l.s=right, label=$U_\nu(x+a\hat{\mu})$}{r0,r1}
 \fmf{fermion,l.d=4.0,l.s=top,   label=$U^\dagger_\mu(x+a\hat{\nu})$}{r1,l1}
\end{fmfgraph*}
}
\end{fmffile}
\end{minipage}
}}
\hspace{7em}
}
\fi

\jname{Annu. Rev. Nucl. Part. Sci.}
\jvol{72}
\jyear{2022}
\doi{10.1146/annurev-nucl-010622-120608}

\begin{document}

% Page header
\markboth{A. Meyer, A. Walker-Loud, C. Wilkinson}{LQCD relevance for the few-GeV neutrino program}

% Title
\title{Status of Lattice QCD Determination of Nucleon Form Factors
 and their Relevance for the Few-GeV Neutrino Program}

%Authors, affiliations address.
\author{Aaron S. Meyer$^{1,2}$,
Andr\'{e} Walker-Loud$^2$,
Callum Wilkinson$^3$
\affil{$^1$Department of Physics, University of California, Berkeley, CA, 94720, USA}
\affil{$^2$Nuclear Science Division, Lawrence Berkeley National Laboratory, Berkeley, CA, 94720, USA}
\affil{$^3$Physics Division, Lawrence Berkeley National Laboratory, Berkeley, CA, 94720, USA}
}

\begin{abstract}
Calculations of neutrino-nucleus cross sections begin with the neutrino-nucleon interaction, making the latter critically important to flagship neutrino oscillation experiments, despite limited measurements with poor statistics. Alternatively, lattice QCD (LQCD) can be used to determine these interactions from the Standard Model with quantifiable theoretical uncertainties. Recent LQCD results of $g_{\mathrm{A}}$ are in excellent agreement with data, and results for the (quasi-)elastic nucleon form factors with full uncertainty budgets are expected within a few years. We review the status of the field and LQCD results for the nucleon axial form factor, $F_{\mathrm{A}}(Q^2)$, a major source of uncertainty in modeling sub-GeV neutrino-nucleon interactions. Results from different LQCD calculations are consistent, but collectively disagree with existing models, with potential implications for current and future neutrino oscillation experiments. We describe a road map to solidify confidence in the LQCD results and discuss future calculations of more complicated processes, important to few-GeV neutrino oscillation experiments.
\end{abstract}

%Keywords, etc.
\begin{keywords}
Neutrino Oscillations, Nucleon Form Factors, Lattice QCD
%keywords, separated by comma, no full stop, lowercase
\end{keywords}

\maketitle

%Table of Contents
\tableofcontents

% ------------------------------------------------------------------------------
% Introduction
\section{Introduction\label{sec:intro}}

A major experimental program is underway which seeks to measure
as of yet unknown properties associated with the change of flavor of neutrinos.
In particular, the neutrino mass hierarchy and%
\begin{marginnote}
\entry{CP}{charge-parity}
\end{marginnote}%
CP violating phase
of neutrinos still remain to be measured, with additional focuses on measuring
oscillation parameters with high precision and testing whether the current
three-flavor mixing paradigm is sufficient~\cite{Esteban:2020cvm, ParticleDataGroup:2020ssz}.
These goals introduce stringent requirements on the precision of current and future experiments.
High-intensity beams are required to produce a sufficient flux of neutrinos to accumulate the necessary statistics.
Increased statistics place additional burden on our understanding of the systematic uncertainties needed for the experimental program.

Two large-scale, next-generation experiments designed to meet these experimental constraints
are
DUNE~\cite{Abi:2020wmh}
and
Hyper-K~\cite{Hyper-Kamiokande:2018ofw}.
DUNE has a broad neutrino energy spectrum with a peak at a neutrino energy of $\approx$2.5 GeV,
with significant contributions between 0.1--10 GeV, over a 1295 km baseline.
Hyper-K has a narrow neutrino energy spectrum peaked at a neutrino energy of $\approx$0.6 GeV, with significant
contributions between 0.1--2 GeV, over a 295 km baseline. Despite their different energies (E) and
baselines (L), both experiments sit at a similar L/E, and therefore probe similar oscillation physics.%
%-------------------------------------------------------------------------------
\begin{marginnote}
    \entry{DUNE}{Deep Underground Neutrino Experiment}
    \entry{Hyper-K}{Hyper Kamiokande}
\end{marginnote}%
%-------------------------------------------------------------------------------
At the few-GeV energies of interest, $\nu N$ interactions have many available interaction channels,
including quasielastic, resonant, and deep inelastic scattering~\cite{zeller12, hayato_review_2014, Mosel:2016cwa, Katori:2016yel, NuSTEC:2017hzk}.

Theoretical models that make different physical assumptions are typically used to model each interaction channel, with {\it ad hoc} interpolations to fill in gaps between the models.
Additionally, all current and planned experiments use target materials predominantly composed of hydrocarbons, liquid argon or water, in which the nucleons are not free. These avoid serious experimental complications associated with using elementary targets (e.g., liquid hydrogen) and increase the interaction rate in a given detector volume due to their higher densities.
However, the presence of multiple interaction channels and the addition of nuclear effects significantly complicates the analysis of data from neutrino experiments and gives rise to a major source of systematic uncertainty:
intranuclear motion can be significant relative to the energy transfer for the interactions of interest;
interactions with correlated nucleon-nucleon states can modify or redistribute the interaction strength;
rescattering of pions and nucleons in the nucleus can confuse the relationship between the particles associated with the primary interaction channels and those observed in the detector;
rescattering can manifest as changes to the particle content and changes to the fraction of energy lost to neutrons, which are typically unobservable in the detectors utilized by few-GeV neutrino oscillation experiments.

A significant challenge impeding progress towards a consistent theoretical description of
$\nu A$ scattering is the lack of data with which to benchmark parts of the calculation.
For example, neutrino quasielastic scattering ($\nu_{l} + n \rightarrow l^{-} + p$ or $\bar{\nu}_{l} + p \rightarrow l^{+} + n$) is the simplest of the relevant hard scattering processes, and dominates the neutrino cross section below energies of $\approx$1 GeV.%
\begin{marginnote}
  \entry{$\nu A$}{Neutrino-nucleus}
  \entry{$\nu N$}{Neutrino-nucleon}
\end{marginnote}%
However, modern experiments using nuclear targets are unable to measure it without significant nuclear effects~\cite{garvey_review_2014, NuSTEC:2017hzk}.
Instead, they select a specific interaction topology, such as one muon and no pions, that will be dominated by quasielastic processes. This event selection will still have significant contributions from resonant pion production events where the pion has rescattered in the nucleus and has either been absorbed or has lost sufficient energy to be below detection threshold.
Given the challenge to benchmark neutrino cross-section models for quasielastic scattering (and other hard-scattering processes) with new $\nu A$ datasets, experimentalists and theorists have relied heavily on sparse data from the 1960--1980's from several bubble chamber experiments that used H$_{2}$ or D$_2$ targets~\cite{zeller12, ParticleDataGroup:2020ssz}.
The small neutrino cross section, and relatively weak (by modern standards) accelerator neutrino beams utilized by these early experiments, mean that the available quasielastic event sample on light targets amounts to a few thousand events~\cite{ANL_Barish_1977, BNL_Fanourakis_1980, BNL_Baker_1981, Kitagaki:1983px, Allasia:1990uy}.%
% FOOTNOTE ---------------------------------------------------------------
\footnote{Some constraints on the axial form factor have been obtained from fits
 to electro pion production data.
The fits are parameterized by a low energy theory that is valid in the chiral
 ($m_\pi\to0$) limit and at small 3-momentum and energy transfer,
 with model-dependent systematics that are significant and not typically quantified.
These data will not be considered in this review.
For more details, we refer the reader to Refs.~\cite{Bernard:1993bq,Bernard:2001rs}.
}
%------------------------------------------------------------------------------
These data do not have sufficient power to constrain theoretical models satisfactorily~\cite{Meyer:2016oeg, Hill:2017wgb}. As a result, there is insufficient information about fundamental $\nu N$ scattering processes on which to build a complete model for $\nu A$ scattering, a substantial limitation that has serious implications for the precision goals of future experiments.

Experimentalists are looking for other ways to access neutrino interactions
with elementary targets for the purpose of disambiguating neutrino cross-section
modeling uncertainties.
Safety considerations make it unlikely that new high-statistics bubble-chamber experiments using
hydrogen or deuterium will be deployed to fill this crucial gap.
An alternative possibility is to use various hydrocarbon targets to subtract the carbon interaction contributions from
the total hydrocarbon event rates, and produce ``on hydrogen'' measurements~\cite{PhysRevD.92.051302, PhysRevD.101.092003, Hamacher-Baumann:2020ogq, DUNE:2021tad, Cai:2021vkc}.
These ideas are promising, but typically rely on kinematic tricks that are only relevant for some channels, and it remains to be seen whether the systematic uncertainty associated with modeling the carbon subtraction can be adequately controlled. Such ideas may also be extended to other compound target materials with hydrogen or deuterium components.

LQCD can be used to determine the free nucleon amplitudes that are otherwise not known at the required precision, without the need for another experiment.
LQCD provides a theoretical method for predicting the free nucleon amplitudes directly from the Standard Model of Particle Physics, with systematically improvable theoretical uncertainties.%
%-------------------------------------------------------------------------------
\begin{marginnote}
\entry{LQCD}{Lattice quantum chromodynamics}
\entry{Axial coupling} {The axial form factor at zero momentum transfer, $g_{\mathrm{A}} = F_{\mathrm{A}}(0)$}
\end{marginnote}%---------------------------------------------------------------
Recently, a LQCD milestone was achieved when the nucleon axial coupling%
%FOOTNOTE --------------------------------------------------------------------
\footnote{Within the LQCD community, the axial coupling
 is commonly referred to as the axial charge.}
%---------------------------------------------------------------------
was determined with a 1\% total uncertainty and a value consistent with experiment~\cite{Chang:2018uxx}.
LQCD can also provide percent- to few percent-level uncertainties for the nucleon quasielastic axial form factor with momentum transfers up to a few-${\rm GeV}^2$.
Similarly, current tension in the neutron magnetic form factor parameterization, which is roughly half the size of the total axial form factor uncertainty, can be resolved with LQCD calculations.
Such results are anticipated in the next year or so with computing power available in the present near-exascale computing era.

Building upon these critical quantities, more challenging computations can provide information about nucleon
resonant and nonresonant contributions to vector and axial-vector matrix elements,
such as the $\Delta$ or Roper resonance channels, pion-production,
inclusive contributions in the shallow inelastic scattering region,
or deep inelastic scattering parton distribution functions.
Additionally, LQCD calculations of two-nucleon response functions would provide crucial information for our theoretical understanding of important two-body currents that are needed for building $\nu A$ cross sections from $\nu N$ amplitudes.

Given the present state of the field, in this review we focus on elastic single-nucleon amplitudes, for which we anticipate the LQCD results will produce impactful results for experimental programs in the next year or two.
We begin in Section~\ref{sec:sof} by surveying the existing status and tensions in the field for the single-nucleon (quasi-) elastic form factors.
Then, in Section~\ref{sec:lqcd}, after providing a high-level introduction to LQCD, we survey existing results of the axial form factor. This includes the role of the%
\begin{marginnote}
 \entry{PCAC} {Partially-conserved} axial current
\end{marginnote}%
PCAC relation in the calculations as well as use of the $z$ expansion for combining the continuum and physical pion mass extrapolations.
In Section~\ref{sec:impact}, we discuss the potential impact of using LQCD determinations of the axial form factor when modeling $\nu A$ cross sections.
In Section~\ref{sec:future}, we comment on the most important improvements to be made in LQCD calculations and we conclude in Section~\ref{sec:conclusions}.

%-------------------------------------------------------------------------------
% State of the field
\section{Status of single nucleon (quasi-) elastic form factors\label{sec:sof}}

For charged-current quasielastic $\nu N$ scattering,
the neutron ($|n\rangle$) to proton ($\langle p|$)
interaction is mediated by a $V-A$ weak current,
 given at the quark level by $\bar{u}\gamma_\mu(1- \gamma_5)d$
 (or its conjugate for proton to neutron).
 The nucleon-level amplitude at four-momentum transfer $Q^2 = -q^2$ is parameterized by
\begin{align}\label{eq:nucleon_ff}
\langle p | V^\mu | n \rangle
    &= \bar{U}_p(p+q) \Big[
        F_1^+(q^2) \gamma^\mu
        +\frac{i}{2M} F_2^+(q^2) \sigma^{\mu\nu} q_\nu
    \Big] U_n(p),
\nonumber\\
\langle p | A^\mu | n \rangle
    &= \bar{U}_p(p+q) \Big[
        F_{\mathrm{A}}^+(q^2) \gamma^\mu \gamma_5
        +\frac{1}{M} F_P^+(q^2) q^\mu \gamma_5
    \Big] U_n(p)\, .
\end{align}
The isovector, vector form factors, $F_1^+$ and $F_2^+$, can be precisely estimated from electron-nucleon scattering data.
Electron-proton and electron-neutron scattering are sensitive to linear combinations of the isoscalar, $F_{1,2}^s$ and isovector, $F_{1,2}^3$ form factors.  After isolating $F_{1,2}^3$, approximate isospin symmetry can be used to relate these $\tau_3$ form factors to the charged $\tau_+$ form factors of Equation~\eqref{eq:nucleon_ff}: in the isospin limit, $\langle p| \bar{u}\, \Gamma u - \bar{d}\, \Gamma d |p\rangle = \langle p| \bar{u}\, \Gamma d |n\rangle$
 for Dirac structure $\Gamma$ and $F_{1,2}^3 = F_{1,2}^+$
 for the isovector Dirac and Pauli form factors.
%------------------------------------------------------------------------------
% proton magnetic FF
\begin{figure}
 \centering
 \includegraphics[width=0.7\textwidth]{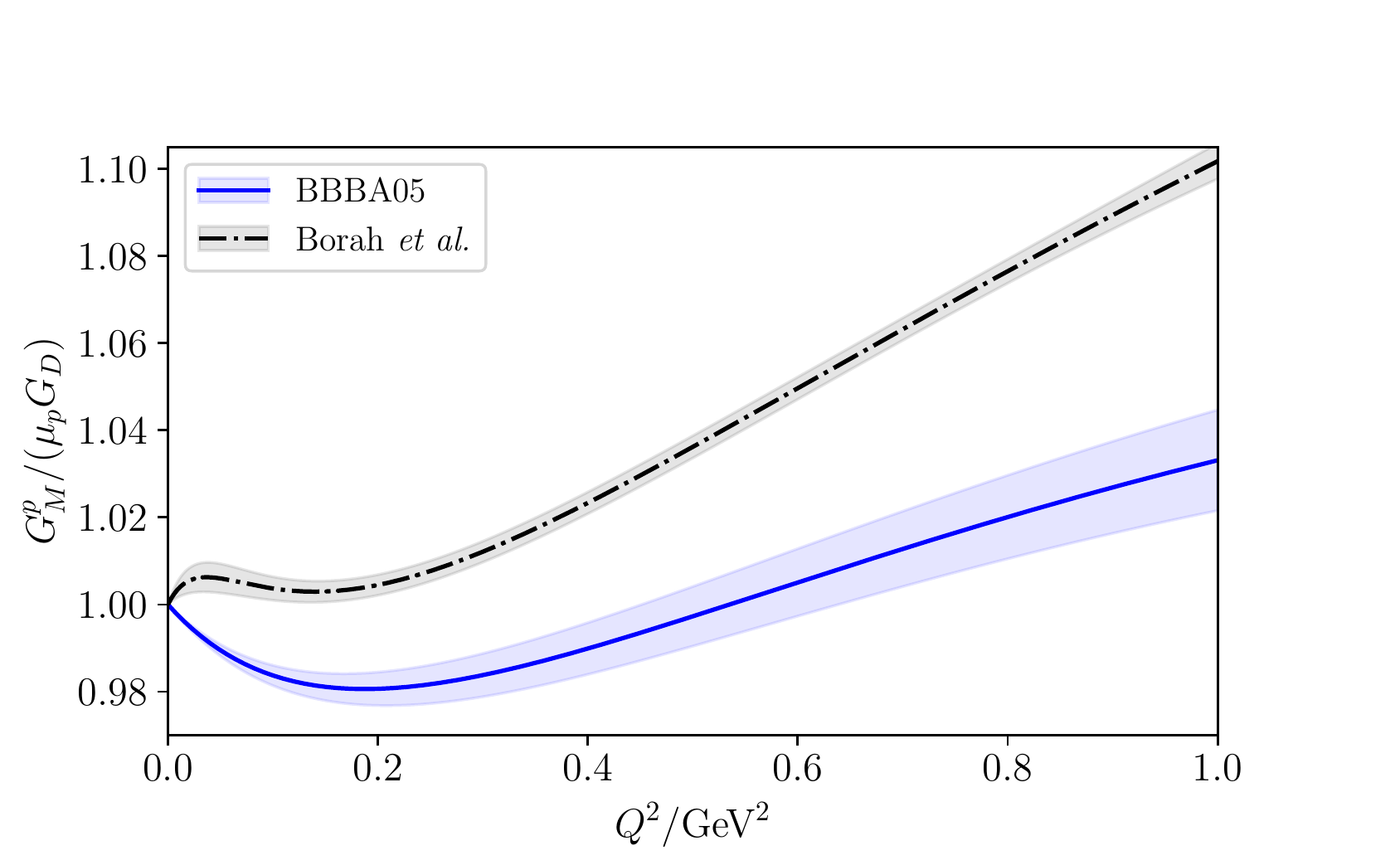}
 \vspace{4pt}
\caption{
Proton magnetic form factor normalized by a reference dipole ansatz
with a dipole mass of $0.84~{\rm GeV}$.
This plot is reproduced from Fig.~4 in Ref.~\cite{Borah:2020gte}
 and the associated supplemental data.
The proton-only fit to a $z$ expansion by Borah {\it et al.}~\cite{Borah:2020gte}
and the BBBA05 parameterization~\cite{Bradford:2006yz} are shown.
\label{fig:protonmagneticff}
}
\end{figure}
%------------------------------------------------------------------------------

However, there is a significant tension in existing parameterizations of the proton magnetic form factor extracted from that data, as
shown in Figure~\ref{fig:protonmagneticff}.
Two different parameterizations of the form factor, normalized by a dipole parameterization, are shown.
The BBBA05~\cite{Bradford:2006yz} are displayed as the lower (blue) band with a solid mean value.%--------------------------------------------------------------
\begin{marginnote}
\entry{dipole parameterization}{$F_D(Q^2)\equiv (1+{Q^2}/{M_D^2})^{-2}$}
\entry{BBBA05}{The Bradford, Bodek, Budd and Arrington 2005 nucleon elastic form factor parameterization~\cite{Bradford:2006yz}}
\end{marginnote}%
A more recent $z$ expansion parameterization from Borah {\it et al.}~\cite{Borah:2020gte} is displayed by the upper (black) band with a dashed mean value.
The tension is significant over all $Q^2 > 0$, at the level of several percent,
including significant disagreement in the slope of the form factor at $Q^2 = 0$.
LQCD computations of the vector form factors are also the most mature of the nucleon matrix elements,
exhibiting no obvious tensions with experimental determinations
at their current level of precision.
A percent-level calculation of the form factor $Q^2$ behavior or a direct calculation of the magnetic form factor slope could potentially discriminate
between the two parameterizations.

The axial coupling
is a key benchmark for LQCD and is precisely known
from neutron decay experiments~\cite{Dubbers:2021wqv}.
LQCD calculations of the axial coupling have historically been low compared to experiment~\cite{Aoki:2021kgd},
 and the discrepancy has been the topic of some controversy.
It is now understood that the treatment of excited state systematics is the main culprit for this discrepancy~\cite{Bar:2017kxh,Ottnad:2020qbw,Aoki:2021kgd}.
This topic, and how it pertains to the full momentum dependence of the form factor,
 will be discussed in detail in Section~\ref{sec:lqcd_pcac}.
With proper control over the excited state contamination, LQCD calculations are now in good agreement with the experimental value~\cite{Jang:2019vkm,Gupta:2018qil,Alexandrou:2020okk,Abramczyk:2019fnf,Park:2021ypf,RQCD:2019jai,Hasan:2019noy,Djukanovic:2021yqg,Harris:2019bih,Liang:2018pis,Shintani:2018ozy,Ishikawa:2018rew}
 with one group achieving a sub-percent determination of $g_{\mathrm{A}}$~\cite{Chang:2018uxx,Berkowitz:2018gqe,Walker-Loud:2019cif}.

The success in calculating the threshold value $g_{\mathrm{A}}$ motivates current efforts
to map out $F_{\mathrm{A}}(Q^2)$ of importance to the long-baseline neutrino program.
The need for this is clear: sparse data from deuterium bubble-chamber experiments do not constrain
the axial form factor precisely.
The popular dipole ansatz has a shape that
is overconstrained by data resulting in an underestimated uncertainty.
Employing a model-independent $z$ expansion parameterization
relaxes the strict shape requirements of the dipole and yields
a more realistic uncertainty that is nearly an order of magnitude larger~\cite{Meyer:2016oeg}.
The axial radius, which is proportional to the slope of the form factor at $Q^2=0$,
has a 50\% uncertainty when estimated from the deuterium scattering data,
or $\approx35\%$ if deuterium scattering and muonic hydrogen are considered
together~\cite{Hill:2017wgb}.
Given that a modern $\nu N$ scattering experiment is extremely unlikely, LQCD is the only viable method to improve our understanding of the axial form factor to the required level of precision.

A striking feature of LQCD calculations of $F_{\mathrm{A}}(Q^2)$
is the slower fall off with increasing $Q^2$ than what is extracted from experiment (Section~\ref{sec:lqcd_results}).
This preference is consistently reproduced by several lattice collaborations using
independent computation methods, lending more credence to the result.
The nucleon cross section is obtained by integrating over $Q^2$, and the slower falloff with $Q^2$
translates to an enhancement
by as much as 30--40\%
for neutrino energies greater than $1~{\rm GeV}$ (Section~\ref{sec:impact}).
In addition, the precision of the axial form factor uncertainty from LQCD
is small enough to be sensitive to the tension between vector form factor parameterizations.

%------------------------------------------------------------------------------
% nu-N cross section
\begin{figure}%[hbt!]
 \centering
 \includegraphics[width=0.7\textwidth]{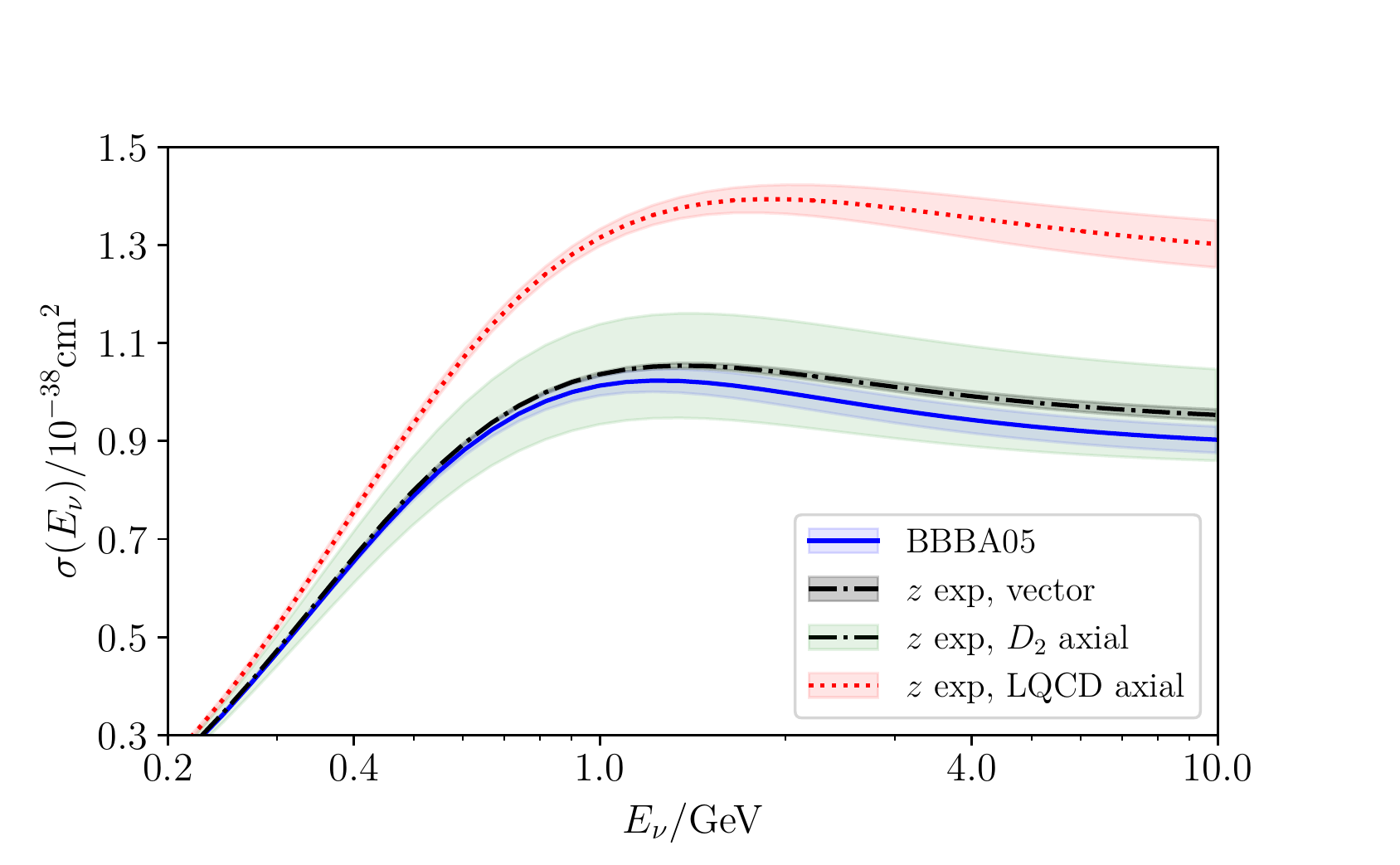}\vspace{4pt}
\caption{
 Neutrino cross sections on a free neutron, with their uncertainty bands,
 for various choices of parameterization explained in the bullet list provided in the Section~\ref{sec:sof}.
 \label{fig:nucleonxsec}
}
\end{figure}

The aforementioned situation with nucleon form factors is depicted in Figure~\ref{fig:nucleonxsec}.
The labels correspond to the following form factor parameterization choices and uncertainties:
\begin{description}
\item[BBBA05] Vector form factors from BBBA05~\cite{Bradford:2006yz},
and axial form factors from Meyer~{\it et al.}~\cite{Meyer:2016oeg}.
The uncertainty is taken only from the BBBA05 parameterization,
with all fit parameters assumed to be uncorrelated.

\item[$z$ exp, vector] Vector form factors from Borah~{\it et al.}~\cite{Borah:2020gte},
and axial form factors from Meyer~{\it et al.}~\cite{Meyer:2016oeg}.
The uncertainty is taken only from Borah~{\it et al.}

\item[$z$ exp, ${\rm D}_{2}$ axial] the same as ``$z$ exp, vector,''
with the uncertainty taken only from Meyer~{\it et al.} instead.

\item[$z$ exp, LQCD axial] Vector form factors from Borah~{\it et al.}~\cite{Borah:2020gte},
and the axial form factor with its uncertainty from an LQCD simulation on a single physical mass ensemble~\cite{Meyer:2021vfq}.
\end{description}
Of particular note is the observed tension between the black and blue bands,
which results from the tension between proton magnetic form factor parameterizations
(see Figure~\ref{fig:protonmagneticff}).
The width of the upper red band, which comes from the LQCD results,
is comparable in size to the discrepancy between these black and blue curves.
The LQCD uncertainty is also noticeably smaller than the green band that arises from the deuterium scattering determination of $F_{\mathrm{A}}(Q^2)$.
The normalization of the red curve is higher due to the slower fall off of the axial form factor.

% ------------------------------------------------------------------------------
% Lattice QCD
\section{LQCD determinations of nucleon form factors\label{sec:lqcd}}

LQCD has been and remains one of the major uses of the world's leadership computing facilities.
There is an extensive literature on LQCD covering the broad range of technical and formal aspects that are necessary to carry out state of the art calculations, for which we can not do justice in this review.
For an in depth introduction to LQCD, we refer readers to the text books~\cite{Smit:2002ug,DeGrand:2006zz,Gattringer:2010zz}. In this review, we provide a high-level summary of general issues that must be addressed as well as issues specific to LQCD calculations of nucleon matrix elements and form factors.
These issues are also discussed in detail in the bi-annual FLAG Reviews, see for example the most recent~\cite{Aoki:2021kgd}.%
%-------------------------------------------------------------------------------
\begin{marginnote}
\entry{FLAG}{Flavour Lattice Averaging Group}
\end{marginnote}

The promise of LQCD is to provide Standard Model predictions of low-energy hadronic and nuclear quantities with fully quantified theoretical uncertainties.
To achieve this goal, several sources of systematic uncertainty must be assessed.
These include extrapolations to the continuum and infinite volume limits as well as an extrapolation or interpolation to the physical quark mass limit.
At least three values of the lattice spacing, $a$, of $\mathrm{O}(a\lesssim0.12\textrm{ fm})$ are required to ascertain if the leading discretization corrections are sufficient or not to describe the observed scaling violations (do all three results lie on a straight line or can one detect higher-order curvature?).
For the finite volume effects, a rule of thumb has been established from experience, that one requires calculations with $m_\pi L \gtrsim4$ (where $L$ is the spatial extent of the lattice volume) in order to keep these finite size corrections at the level of $\lesssim1-2\%$ and at least qualitatively described by the leading analytic formulae.%
\begin{marginnote}
    \entry{$\chi$PT}{Chiral Perturbation Theory: the low-energy effective field theory of QCD}
\end{marginnote}%
For the light-quark mass dependence, $\chi$PT may be able to guide the extrapolations.
However, for the nucleon, the convergence of $\chi$PT is not yet established, even at the physical pion mass with evidence of lack of convergence for the nucleon mass and $g_{\mathrm{A}}$~\cite{Chang:2018uxx,Walker-Loud:2019cif}.
As we will discuss more in Section~\ref{sec:calc_anatomy}, there are two additional significant sources of uncertainty for nucleons which are the exponentially%
\begin{marginnote}
    \entry{S/N}{Signal-to-noise}
\end{marginnote}%
degrading S/N problem and excited state contamination.

%-------------------------------------------------------------------------------
% LQCD Intro
\subsection{LQCD: a high level summary}
The QCD path integral is quadratic in the quark fields allowing for an analytic integration over the fermionic fields. In Euclidean space, one has the gluonic integral
\begin{equation}\label{eq:Z_QCD}
Z_{\mathrm{QCD}} = \int D U\, {\rm Det}[\Dslash(U) + m_q]\, e^{-S_{\mathrm{G}}(U)},
\end{equation}
with gluon action $S_{\mathrm{G}}(U)$ and the determinant of the quark operator ${\rm Det}[\Dslash(U) + m_q]$, for each flavor of quark simulated.
Even at finite lattice spacing and volume, the multi-dimensional integral is vastly too large to perform.
However, in Euclidean space, both $S_{\mathrm{G}}$ and the fermion determinant are real and positive for zero chemical potential, and so the integral can be approximated with an HMC algorithm~\cite{Duane:1987de} using the factor ${\rm Det}[\Dslash(U) + m_q]\, e^{-S_{\mathrm{G}}(U)}$ as the importance sampling weight.%
%-------------------------------------------------------------------------------
\begin{marginnote}
\entry{HMC}{Hybrid Monte Carlo}
\entry{Configurations}{Samples of the gluon field}
\entry{Ensemble}{A set of configurations all generated with the same bare QCD parameters}
\end{marginnote}%
%-------------------------------------------------------------------------------
In this way, a large number of configurations of gauge fields can be generated, providing a stochastic determination of the correlation functions
\begin{equation}
\langle O \rangle = \frac{1}{N_{\rm cfg}}\sum_{i=1}^{N_{\rm cfg}} O[U_i]
    +\mathrm{O}\left(\frac{1}{\sqrt{N_{\rm cfg}}} \right)\, ,
\end{equation}
where $O[U_i]$ is the correlation function evaluated on configuration $i$.
The most expensive part of generating the configurations is evaluating the fermion determinant for the light and strange quarks.
This is done with the use of pseudo-fermions (bosonic fields, $\phi$)
\begin{equation}
Z_\psi = \int D\bar{\psi}D\psi\, e^{-\bar{\psi}[\Dslashe[U]+m_q]\psi}
    = {\rm Det}[\Dslash(U) + m_q]
    = \int D\phi^\dagger D\phi\, e^{-\phi^\dagger \frac{1}{\Dslashe[U]+m_q} \phi}
\end{equation}
for which the bilinear operator is the inverse of the Dirac operator, which is a large, sparse matrix.
Most of the algorithmic development for accelerating LQCD has gone into efficiently solving these large sparse matrices with large condition numbers.  In particular, this is a problem very well suited for GPUs
for which we have an advanced library, QUDA~\cite{Clark:2009wm,Babich:2011np}, developed for the international community.%
%-------------------------------------------------------------------------------
\begin{marginnote}
\entry{GPU}{Graphical Processing Unit}
\end{marginnote}%
%-------------------------------------------------------------------------------

There are many valid choices one can make in constructing the discretized lattice action, provided continuum QCD is recovered as $a\rightarrow0$.
This is known as the universality of the continuum limit, with each choice only varying at finite lattice spacing.
Deviations from QCD, which arise at finite $a$, are often called \textit{discretization corrections} or \textit{scaling violations}.%
%-------------------------------------------------------------------------------
\begin{marginnote}
\entry{Universality}{All valid choices of discretized QCD become QCD as $a\rightarrow0$}
\entry{EFT}{Effective field theory}
\end{marginnote}%
%-------------------------------------------------------------------------------
Universality is a property that can be proved in perturbation theory but must be established numerically given the non-perturbative nature of QCD.  For sufficiently small lattice spacings, one can use EFT to construct a continuum theory that encodes the discretization effects in a tower of higher dimensional operators. This is known as the Symanzik EFT for lattice actions~\cite{Symanzik:1983dc,Symanzik:1983gh}.
One interesting example involves the violation of Lorentz symmetry at finite lattice spacing: in the Symanzik EFT, the operators that encode Lorentz violation scale as $a^2$ with respect to the operators that survive the continuum limit. Thus, Lorentz symmetry is an accidental symmetry of the continuum limit.  It is not respected at any finite lattice spacing, but the measurable consequences vanish as $a^2$ for sufficiently small lattice spacing.

As a concrete
example of the Symanzik EFT, consider the discretized gluon action.
The link fields are Wilson lines
\begin{equation}
U_\mu(x) = \exp\left\{i a\int_0^1 dt A_\mu(x +(1-t)a\hat{\mu}) \right\}
    \approx \exp\left\{i a \bar{A}_\mu(x) \right\}\, .
\end{equation}
The gluon field $A_\mu(x)$ can be approximated as constant over the interval $[x, x+a\hat{\mu}]$, as expressed by $\bar{A}_\mu(x)$, with $a$ being the lattice spacing.
This parameterization allows for the construction of a discretized theory that preserves gauge-invariance~\cite{Wilson:1974sk}, a key property of gauge theories.
In the continuum, the gluon action-density is given by the product of field strength tensors, which are gauge-covariant curls of the gauge potential.
When constructing the discretized gluon-action, it is therefore natural to use objects which encode this curl of the gauge potential.  The simplest such object is referred to as a ``plaquette'' and given by
\begin{equation}
  \Umunu
    =U_{\mu\nu}(x)
    =U_\mu(x)U_\nu(x+a\hat{\mu}) U^\dagger_\mu(x+a\hat{\nu}) U^\dagger_\nu(x)\, .
\end{equation}
For small lattice spacing, this Wilson gauge-action reduces to the continuum action plus irrelevant (higher dimensional) operators which vanish in the continuum limit%
%FOOTNOTE
\footnote{In the renormalization sense, these are operators of mass dimension $[O]>4$, such that their dimensionful couplings scale as $a^n=\Lambda^{-n}$ where $\Lambda$ is a high energy scale
that goes to infinite in the continuum limit for LQCD, and $n=[O]-4$}
%--------------------------------------------------------------
\begin{align}\label{eq:gluon_action}
S_{\mathrm{G}}(U) &= \beta \sum_{n=x/a} \sum_{\mu<\nu}
    \textrm{Re}\left[ 1 - \frac{1}{N_c} \textrm{Tr} \left[U_{\mu\nu}(n) \right]\right]
\nonumber\\&=
    \frac{\beta}{2N_c}
    a^4 \sum_{n=x/a,\mu,\nu}
    \left[
        \frac{1}{2} \textrm{Tr} \left[ G_{\mu\nu}(n)G_{\mu\nu}(n)\right]
        +\mathrm{O}(a^2)
    \right]\, ,
    & \rightarrow \beta = \frac{2N_c}{g^2}\, .
\end{align}
The continuum limit, which is the asymptotically large $Q^2$ region, is therefore approached as $\beta\rightarrow\infty$ where $g(Q^2)\rightarrow 0$.%

The inclusion of quark fields adds more variety of lattice actions, each with their own benefits and drawbacks.
There are four commonly used fermion discretization schemes which are known as staggered fermions~\cite{Kogut:1974ag,Banks:1975gq,Banks:1976ia,Susskind:1976jm}, clover-Wilson fermions~\cite{Sheikholeslami:1985ij}, twisted mass fermions~\cite{Frezzotti:2000nk} and DWF~\cite{Kaplan:1992bt,Shamir:1993zy,Furman:1994ky}.%
%-------------------------------------------------------------------------------
\begin{marginnote}
\entry{DWF}{Domain Wall Fermions}
\end{marginnote}%
%-------------------------------------------------------------------------------
In this review, we comment that:
\begin{itemize}[leftmargin=*]
\item Staggered fermions are the least expensive to simulate numerically, have leading scaling violations of $\mathrm{O}(a^2)$, and they have a remnant chiral symmetry protecting the quark mass from additive mass renormalization.  However, they split the four components of the fermion spinor onto different components of a local hypercube, mixing the Dirac algebra with spacetime translations.  This significantly complicates their use for baryons~\cite{Golterman:1984dn,Bailey:2006zn,Lin:2019pia}.

\item Clover-Wilson fermions are the most commonly used discretization scheme given their theoretical simplicity and preservation of all symmetries except chiral symmetry.  The explicit breaking of chiral symmetry with the Wilson operator means the light quark masses must be finely tuned against ultraviolet chiral symmetry breaking that scales as $1/a$, after which there remain residual $\mathrm{O}(a)$ chiral symmetry breaking effects.  It is well known, albeit laborious, how to non-perturbatively remove these leading $\mathrm{O}(a)$ scaling violations~\cite{Luscher:1996sc,Luscher:1996ug,Luscher:1996jn,Capitani:1998mq}, which must be done for both the action as well as matrix elements.

\item Twisted mass fermions are a variant of Wilson fermions that exploits the approximate $SU(2)$ chiral symmetry of QCD to introduce a twisted quark mass term, $i\mu\g_5 \tau_3$.
This term is used to automatically remove the leading $\mathrm{O}(a)$ discretization effects~\cite{Frezzotti:2003ni}%
 , a benefit generically referred to as $\mathrm{O}(a)$ improvement,
 at the expense of introducing isospin breaking at finite lattice spacing.

\item The fourth most common discretization are DWF, which introduce a fifth dimension to the theory with unit links (the gluons are not dynamic in the fifth dimension) with the left and right handed fermions bound to opposite sides of the fifth dimension of size $L_5$.  The overlap of these left and right modes gives rise to an explicit chiral symmetry breaking that is exponentially suppressed by the extent of the fifth dimension.  For sufficiently small chiral symmetry breaking (large $L_5$), DWF are also automatically $\mathrm{O}(a)$ improved.
While very desirable, DWF are more expensive to simulate numerically, both because of the extra fifth dimension and also because the algorithmic speed up offered by multi-grid computational technique, which works tremendously for clover-Wilson fermions on GPUs~\cite{Clark:2016rdz} but is not yet fleshed out for DWF~\cite{Boyle:2014rwa,Cohen:2011ivh,Yamaguchi:2016kop,Brower:2020xmc,Boyle:2021wcf}.

\item A final common variant of action is one in which the fermion discretization used in the generation of the gauge fields (the sea quarks) and the action used when generating quark propagators (the valence quarks) are different: this is known as a \textit{mixed action}~\cite{Renner:2004ck}.
The most common reason to use such an action is to take advantage of numerically less expensive methods to generate the configurations while retaining good chiral symmetry properties of the valence quarks, which is known to suppress chiral symmetry breaking effects from the sea-quarks~\cite{Bar:2002nr,Bar:2005tu,Tiburzi:2005is,Chen:2007ug}.

\end{itemize}
As mentioned above, a key assumption of LQCD is that all varieties of lattice action, for sufficiently small lattice spacing, are approximated by continuum QCD plus irrelevant operators whose contributions vanish in the continuum limit.
It is important for the field to test this assumption of universality by computing the same quantities with a variety of lattice actions, both at the level of gluons as well as the fermions, in order to gain confidence in the results that are extrapolated to the physical point.

% ------------------------------------------------------------------------------
% anatomy of LQCD calculation
\subsection{Anatomy of LQCD calculations of nucleon form factors\label{sec:calc_anatomy}}

Hadron masses are determined from LQCD by constructing two-point correlation functions in a mixed time-momentum representation.
A common strategy uses spatially local creation operators (sources) and momentum space annihilation operators (sinks), taking advantage of momentum conservation to select the source momentum.
The non-perturbative nature of QCD means we do not know how to construct the nucleon wave function, and so we utilize \textit{interpolating operators} which have the quantum numbers of the state we are interested in.
These creation and annihilation operators will couple to all eigenstates of QCD with the same quantum numbers, giving rise to a two-point function with a spectral decomposition
\begin{align}\label{eq:2pt}
    C(t,\mathbf{p}) &= \sum_{\mathbf{x}} e^{-i \mathbf{p\dotp x}}
        \langle \O| O(t,\mathbf{x}) O^\dagger(0,\mathbf{0}) | \O \rangle
    =
    \sum_{n=0}^\infty z_n(\mathbf{p}) z_n^\dagger(\mathbf{p}) e^{-E_n(\mathbf{p})t}\, .
\end{align}
In this expression, $|\O\rangle$ is the vacuum state,
$z_n(\mathbf{p}) = \sum_{\mathbf{x}}e^{-i\mathbf{p\dotp x}} \langle \O|O(0,\mathbf{x})|n\rangle$
and $z_n^\dagger(\mathbf{p}) = \langle n(\mathbf{p})|O^\dagger(0,\mathbf{0})|\O\rangle$.
To go from the first equality to the second, we have inserted a complete set of states, $1=\sum_n |n\rangle\langle n|$ and we have used the time-evolution operator to shift the annihilation operator to $t=0$ and expose the explicit time dependence.%
% FOOTNOTE --------------------------------------------------------------
\footnote{The Hamiltonian, $\hat{H}$, is used to time evolve the operator $O(t,\mathbf{x}) = e^{\hat{H}t} O(0,\mathbf{x}) e^{-\hat{H}t}$.}
%------------------------------------------------------------------------------
As we will discuss in more detail below, it is more desirable to instead build both source and sink operators
in momentum space.
Momentum space sources are not commonly used as they are significantly more numerically expensive to generate.

For large Euclidean time, the correlation function will be dominated by the ground state as the excited states will be exponentially suppressed by the energy gap
\begin{align}
&C(t) = z_0 z_0^\dagger e^{-E_0 t}\left[
        1 + r_1 r^\dagger_1 e^{-\Delta_{1,0}t} + \cdots \right]\, ,&
&\D_{m,n}= E_m - E_n\, ,&
&r_n = \frac{z_n}{z_0}\, .&
\end{align}
It is useful to construct an \textit{effective mass} to visualize at which time $t$, the ground state begins to saturate the correlation function
\begin{align}
m_{\rm eff}(t) &= \ln \left( \frac{C(t)}{C(t+1)} \right)
    =
    E_0 + \ln\left( 1 + \sum_{n=1} r_n r^\dagger_n e^{-\Delta_{n,0}t}\right)\, .
\end{align}

For nucleon two-point functions, the S/N ratio degrades exponentially at large Euclidean time~\cite{Lepage:1989hd}
\begin{equation}
\lim_{{\rm large}\ t} {\rm S/N}
    \propto \sqrt{N_{\mathrm{sample}}} e^{-(m_{\mathrm{N}} - \frac{3}{2}m_\pi)t}\, .
\end{equation}
In the region in time when the ground state begins to saturate the correlation functions, typically around $t\approx 1$~fm, the noise becomes significant, which makes the correlation functions in this region susceptible to correlated fluctuations that can bias a simplistic single-state analysis.
This forces measurements to be made at Euclidean times where
 excited state contamination is still appreciable,
 which adds an extra source of systematic uncertainty.
As the pion mass is reduced towards its physical value, the energies of the excited states decrease, as the lowest lying excited state is typically a nucleon-pion in a relative $P$-wave.
At the same time, the energy scale that governs the exponential degradation of the signal also grows.
The former issue means calculations must be performed at larger Euclidean time to suppress the slowly decaying excited states, and the latter issue means we need exponentially more statistics to obtain a fixed relative uncertainty at a given Euclidean time.

The most common method of constructing three-point correlation functions follows a strategy similar to the two-point correlation functions, beginning with spatially local sources.
A nucleon three-point function with current $j_\G$ is constructed with interpolating operators $N(\tsep,\mathbf{x})$ and $N^\dagger(0,\mathbf{0})$,
\begin{align}
C_\G(\tsep,\t) &= \sum_{\mathbf{x,y}}e^{-i\mathbf{p\dotp x} +i\mathbf{q\dotp y}}
    \langle\O|N(\tsep,\mathbf{x}) j_\G(\t,\mathbf{y}) N^\dagger(0,\mathbf{0}) |\O\rangle
\nonumber\\&=
    \sum_{\mathbf{x,y}}e^{-i\mathbf{p\dotp x} +i\mathbf{q\dotp y}}
    e^{-E_n(\tsep-\t)}e^{-E_m\t}
    \langle\O|N(0,\mathbf{x})|n\rangle\langle n|j_\G(0,\mathbf{y})|m\rangle\langle m| N^\dagger(0,\mathbf{0}) |\O\rangle
\nonumber\\&=
    \sum_{n,m} z_n(\mathbf{p})z_m^\dagger(\mathbf{p-q})e^{-E_n(\tsep-\t)}e^{-E_m \t} g_{n,m}^\G(\mathbf{q})\, ,
\end{align}
where $g_{n,m}^\G$ are the matrix elements of interest, and in principle, all other quantities can be determined from two-point functions, a point we will return to.
Often, the sink is projected to zero momentum, $\mathbf{p}=0$, and momentum conservation selects an incoming state with momentum $-\mathbf{q}$.%
%-------------------------------------------------------------------------------
\begin{marginnote}
\entry{$j_\G$}{quark bilinear currents of Dirac structure $\G$ and unspecified flavor structure $\bar{q}\, \G\, q$}
\entry{$g_{n,m}^\G$}{hadronic matrix elements of interest from state $m$ to $n$ with implicit momentum and energy dependence, $\langle n| j_\G |m\rangle$}
\entry{$\tsep$}{The time-separation between the sink and source}
\end{marginnote}%
%-------------------------------------------------------------------------------
A typical calculation is performed with a sequential propagator~\cite{Martinelli:1988rr}
whose source is obtained by taking forward propagators from the origin and contracting the spin and color indices for all but one quark operator.
In the case of the nucleon three-point correlation function,
 the sequential propagator could originate either from the current insertion or from the sink.%
% FOOTNOTE ---------------------------------------------------------------------
\footnote{There are alternative methods for computing nucleon structure known as the one-end trick~\cite{Foster:1998vw,McNeile:2006bz,Alexandrou:2013xon} and a variant of the Feynman-Hellmann Theorem~\cite{CSSM:2014uyt}, but these are not in wide use.}
%-------------------------------------------------------------------------------

The most significant challenge for determining the nucleon matrix elements and subsequent form factors is dealing with the excited state contamination, an issue that is compounded by the degrading S/N.
If the nucleon two-point function is becoming saturated by the ground state at $t\approx1$~fm, ideally three-point functions would use values of $\tsep\gtrsim 2$~fm.
However, the S/N ratio of the three-point functions decays more rapidly than the two-point functions.
In practice, a few values of $\tsep$ in the range $0.8 \lesssim \tsep \lesssim 1.5$~fm are used and a fit to the time dependence is used to isolate the ground state matrix elements.
Fits that allow for just one excited state require three values of $\tsep$ to not be overconstrained.
Most results have been generated with three or fewer values of $\tsep$.
Ref.~\cite{Chang:2018uxx} utilized many values of $\tsep$ to determine $g_{\mathrm{A}}$, and a few other groups have begun advocating for the use of many values of $\tsep$, including small values, to improve control over the excited state contamination~\cite{Hasan:2019noy,Alexandrou:2019brg,He:2021yvm}.
Ref.~\cite{He:2021yvm} utilized 13 values of $\tsep$, which allowed for a systematic study of the uncertainty associated with the truncation of $\tsep$ and a fit to a 5-state model.

At non-zero momentum, the trade-off between excited states and S/N becomes more problematic.
While the energy associated with the signal grows with the momentum, the energy scale associated with the noise is independent of the momentum. Their difference, which is the energy scale that governs the decay of the S/N, therefore grows with increasing momentum.
For values of $Q\gtrsim 2$~GeV, the noise becomes unmanageable unless one uses a smearing profile that couples more strongly to boosted nucleons~\cite{Bali:2016lva}.
Parity is also no longer a good quantum number for boosted nucleons and so the matrix elements couple to both even and odd parity states.  Such contamination can be handled through a variational method that incorporates even and odd parity nucleons~\cite{Stokes:2013fgw,Stokes:2018emx,Stokes:2019zdd}.

Recent results have uncovered some additional aspects of excited state contamination:
\begin{itemize}[leftmargin=*]
\item Nucleon two-point functions constructed as in Equation~\eqref{eq:2pt} are insufficient to reliably determine any of the excited states.  Different choices of $\tmin$ and different reasonable priors in a Bayesian analysis support excited states that differ by several sigma, also resulting in sensitivity of the ground state spectrum~\cite{Park:2021ypf,He:2021yvm} and matrix elements~\cite{Jang:2019vkm,Gupta:2021ahb}.

\item In contrast, the curvature in the three-point functions associated with the current insertion time provides extra constraints that make the determination of the spectrum stable and robust while varying $\tmin$, the number of excited states, the excited state model and the excited state priors~\cite{He:2021yvm}.
While it is encouraging that the spectrum and matrix elements become stable, such an analysis offers no insight into the nature of the excited states.  For example, with typical values of $m_\pi L\approx4$, the $P$-wave $N(\mathbf{p})\pi(-\mathbf{p})$ excited state energy is essentially the same as the $N\pi\pi$ threshold state.%
\begin{marginnote}
 \entry{$N\pi$}{Nucleon-pion}
 \entry{$N\pi\pi$}{Nucleon and two pion}
\end{marginnote}

\item Many groups determine the spectrum and overlap factors from fits to the two-point functions, and then pass these results into the three-point function analysis without allowing the values to adjust to the global minimum.  Such a choice can either lead to an overestimate of the uncertainty of the three-point functions (if one uses the variability of the spectrum mentioned above), or a biased extraction of the ground state matrix elements (if one uses the ``wrong'' value of the spectrum).  Given the computational setup described above, the robust choice is to perform a global analysis of the two- and three-point functions simultaneously.

\item For zero-momentum transfer,
 where the spectrum of the in and the out states have the same value,
 the use of analysis techniques such as the summation method~\cite{Maiani:1987by}
 or a variant of the Feynman-Hellman
 method~\cite{deDivitiis:2012vs,Bouchard:2016heu} can significantly suppress the excited states~\cite{Capitani:2012gj,He:2021yvm}.
For non-zero momentum transfer, only the Breit-Frame where the momentum of the in and out states is equal and opposite, is amenable to this alternative method~\cite{Gambhir:2019pvw}.

\item The excited state contamination is particularly relevant for the PCAC relation, which we discuss in more detail in Section~\ref{sec:lqcd_pcac}.

\end{itemize}

% ------------------------------------------------------------------------------
% LQCD PCAC
\subsection{Role of PCAC in LQCD results of $F_{\mathrm{A}}(Q^2)$\label{sec:lqcd_pcac}}

Given the challenges in identifying all the sources of systematic uncertainty in the calculation of $g_{\mathrm{A}}$, particularly the excited state contamination, it is prudent to perform cross checks of observables that test for consistency of the results.
The validity of the PCAC relation,
\begin{align}
 \partial^\mu A^{a}_{\mu}(x) = 2 m_q P^{a}(x),
 \label{eq:pcac}
\end{align}
provides a complex consistency check at nonzero $Q^2$.
Here, $A^{a}_\mu$ and $P^{a}$ are the axial and pseudoscalar currents,
 respectively, and $m_q$ is the light quark mass.
The PCAC relation is an exact symmetry in the continuum limit.
When applied to the axial current matrix element of the nucleon, it yields the Goldberger-Treiman relation, which is usually expressed in the $Q^2\approx m_\pi^2$ region.  For arbitrary $Q^2$, it is sometimes referred to as the GGT%
 \begin{marginnote}
 \entry{GGT}{Generalized Goldberger-Treiman relation (Equation~\eqref{eq:ggt})}
 \end{marginnote}%
 relation,
\begin{align}
 2 m_{\mathrm{N}} F_{\mathrm{A}}(Q^2) -\frac{Q^2}{2m_{\mathrm{N}}} \widetilde{F}_{\mathrm{P}}(Q^2) = 2 m_q F_{\mathrm{P}}(Q^2),
 \label{eq:ggt}
\end{align}
 which provides orthogonal checks of individual matrix elements
 for the axial and pseudoscalar currents.
The axial, induced pseudoscalar, and pseudoscalar form factors of the nucleon
 ($F_{\mathrm{A}}$, $\widetilde{F}_{\mathrm{P}}$, and $F_{\mathrm{P}}$, respectively) appear in this expression,
 and $m_{\mathrm{N}}$ is the nucleon mass.
The PPD ansatz, which is only approximate even in the continuum limit,%
\begin{marginnote}
 \entry{PPD}{Pion pole dominance}
 \end{marginnote}%
\begin{align}
 \widetilde{F}^{\rm PPD}_{\mathrm{P}}(Q^2) = \frac{4m_{\mathrm{N}}^2}{Q^2+m_\pi^2} F_{\mathrm{A}}(Q^2),
 \label{eq:ppd}
\end{align}
is obtained
 by carefully considering the leading asymptotic behavior of the
 form factors in the double limit $Q^2\to0$ and $m_q\to0$~\cite{Sasaki:2007gw}.

Initial calculations targeting the axial form factor verified the PCAC relation
 for the full correlation functions but found significant \emph{apparent} violations
 of GGT~\cite{Ishikawa:2018rew,Gupta:2017dwj,Bali:2018qus}. The resolution of this apparent violation
 is now informed by baryon $\chi$PT, which suggests that chiral
 and excited state corrections to the spatial axial, temporal axial, and induced pseudoscalar
 are functionally different and not properly removed.
The axial form factor contributions are largely dominated by loop-level $N\pi$ excited states
 with a highly suppressed tree-level correction.
The correction to the axial form factor is nearly independent of $Q^2$.
On the other hand, corrections to the induced pseudoscalar form factor are
 driven by the tree-level correction and has
 a strong $Q^2$ dependence~\cite{Bar:2018xyi}, with the largest correction at low $Q^2$.
The $N\pi$ loop contribution in the induced pseudoscalar is highly suppressed by
 an approximate cancellation.
The contamination to the pseudoscalar current is redundant with the
 chiral corrections to the axial and induced pseudoscalar form factors and can be obtained
 by application of the PCAC relation.

Scrutiny of the LQCD data has demonstrated many of the features
 predicted by chiral perturbation theory.
$N\pi$ states were initially expected to be negligible due to a volume suppression
 of the state overlap, which makes them invisible to the two-point functions~\cite{Bar:2016uoj}.
However, the three-point axial matrix element can enhance these contributions relative
 to the ground state nucleon matrix element, which is enough to overcome the volume suppression.
The main excited states that contaminate the temporal axial form factor matrix elements
 were shown to be driven by two specific $N\pi$ states,
 characterized by a transition
 of the nucleon state to an $N\pi$ excited state through the axial current or vice versa~\cite{Jang:2019vkm}.
In the language of $\chi$PT, these states contribute to tree-level nucleon-pion graphs with fixed relative momentum~\cite{Bar:2018xyi}.
In contrast to the temporal axial insertion, the spatial axial insertion
 is instead expected to be more strongly affected by a tower of loop-level
 $N\pi$ corrections~\cite{Bar:2018xyi}.
Taking into account these observations,
 analyses that fix the spectrum using the two-point functions alone
 will often miss the important $N\pi$ contamination to the
 axial matrix element~\cite{Jang:2019vkm,He:2021yvm}.

Nucleon matrix elements of the temporal axial current have the largest visible excited state contamination~\cite{Jang:2019vkm,RQCD:2019jai}, which can be at least qualitatively understood with $\chi$PT~\cite{Bar:2018xyi}.
This has led to new analysis strategies that more carefully deal with the $N\pi$ excited state with pion-pole contributions and additionally use the temporal axial current correlator to determine excited states.
These new strategies yield minimal violations of PCAC and an improved understanding of excited state contamination on matrix elements sensitive to $F_{\rm A}(Q^2)$ or $\tilde{F}_{\rm P}(Q^2)$.
Previously, these matrix elements exhibited deviations as large as $40\%$ from the PPD or GGT
 relations at low $Q^2$ where they were expected to work best~\cite{Bali:2014nma,Gupta:2017dwj}.

These results are very encouraging,
 but must be reproduced by several independent LQCD calculations for validation.
At the time of writing,
 not all of the results are free from imposed theoretical expectations.
In order to achieve a more stringent validation, the calculations would have to generally be performed where the correlation functions are saturated by the ground state.
This could be accomplished by taking the correlation function at large Euclidean times,
 but the noise is exponentially larger.
Given the extreme cost of using such a strategy, a better alternative would be to implement a variational method that allows for the use of multi-hadron operators that can explicitly remove the $N\pi$ excited states through a diagonalization of the correlation functions~\cite{Blossier:2009kd}.  We will return to this point in Section~\ref{sec:future}.

% ------------------------------------------------------------------------------
% LQCD results
\subsection{Survey of LQCD results of $F_{\mathrm{A}}(Q^2)$\label{sec:lqcd_results}}

\begin{figure}[hbt!]
\centering
\includegraphics[width=0.75\textwidth]{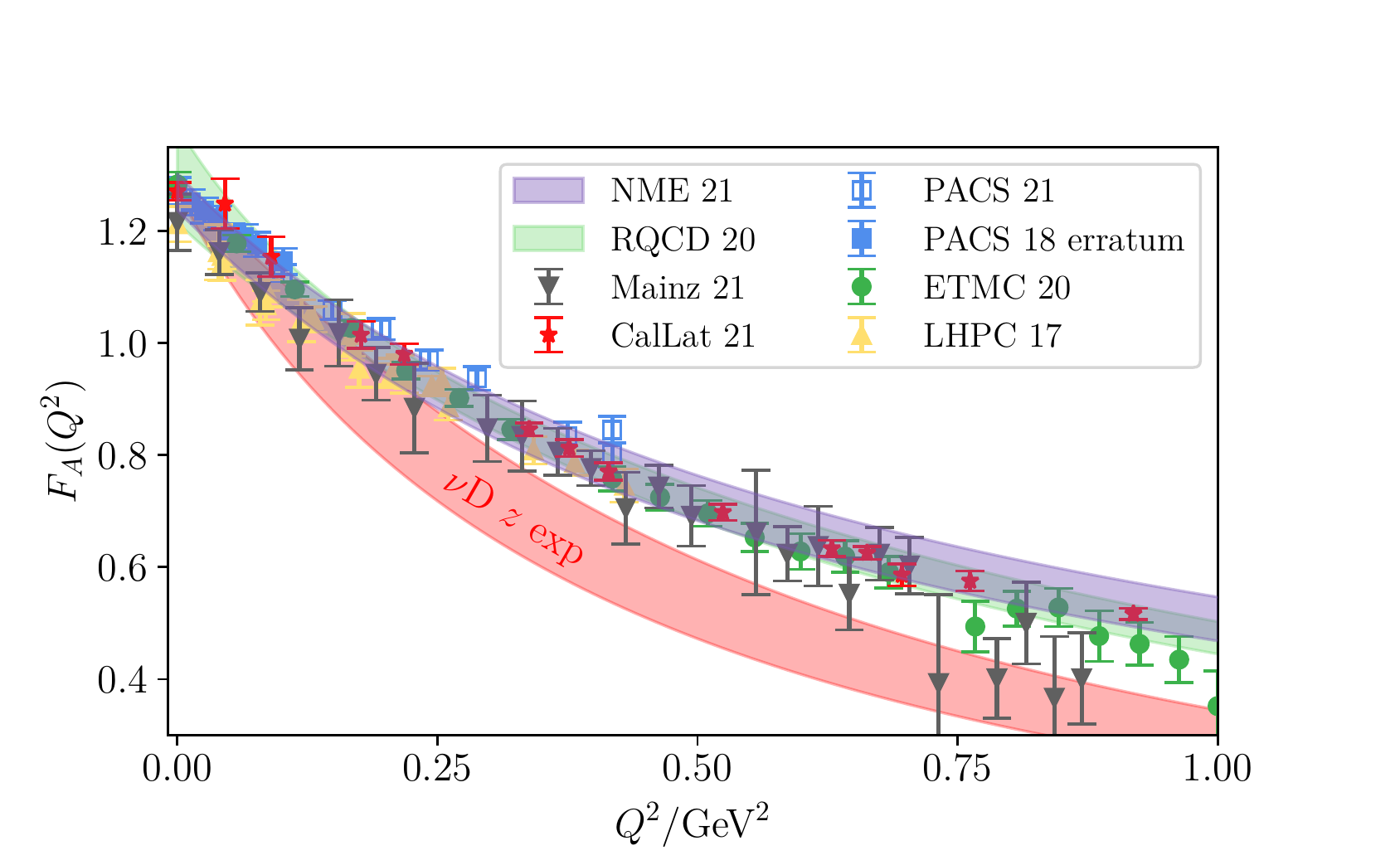}
\vspace{10pt}
\caption{
Published results for the axial form factor at the physical pion mass obtained from LQCD,
 compared with the deuterium extraction from Ref.~\cite{Meyer:2016oeg}.
Results taken from only a single ensemble are plotted as scatter points.
These single-ensemble results will have small but unknown corrections due to chiral, continuum,
 and finite volume systematic shifts.
The NME~\cite{Park:2021ypf} and RQCD~\cite{RQCD:2019jai}
 results are both obtained from fits to several ensembles.
The RQCD perform the full chiral-continuum and finite volume extrapolations to the data,
 fitting to each of the form factors independently for each ensemble but providing
 the constraint that the form factors must satisfy the GGT relation in the continuum.
The NME collaboration also performed a chiral-continuum and finite volume extrapolation
 on their data, but
 their results are based a fit to their five largest volume ensembles neglecting
 effects from lattice spacing, finite volume, and pion mass.
The plotted NME result is obtained
 by inflating the uncertainty on $g_A$ and $b_0$ in Equation 55
 of Ref.~\cite{Park:2021ypf} by a factor of 3
 to account for possible variation due to lattice spacing and quark mass.
}
\label{fig:gaq2_overlay}
\end{figure}

The main deliverables from LQCD calculations of the axial form factor
 are the axial and induced pseudoscalar form factors taken in the continuum,
 infinite volume and physical pion mass limits, complete with a set of parameterization coefficients
 and covariance matrix.
Though the axial coupling
 and radius are useful for connecting with low-energy applications,
 such as electro pion production and neutron decay,
 the needs of neutrino physics in the few-GeV energy range depend
 on the full momentum transfer dependence of the form factor.
Despite the agreement between LQCD and experiment for the axial radius,
 LQCD predicts an axial form factor that falls off more slowly as a function of $Q^2$ than experiment,
 which is shown in Figure~\ref{fig:gaq2_overlay}.
If the axial radius were the only parameter determining the form factor $Q^2$ dependence,
 then these two observations are incompatible.

Though the form factor shape, especially when allowed to explore its full uncertainty,
 is decidedly \emph{not} a dipole, the central value curve determined
 from experiment appears to be dipole-like.
Restricting to dipole shape only, agreement with the axial radius at $Q^2=0$
 would mean the axial form factor should agree over the entire relevant $Q^2$ range,
 a statement that is not supported by the LQCD data.
There is no reason to expect nature to prefer a dipole-like parameterization ---
 the experimental preference toward a dipole-like parameterization is based on
 a few datasets from the 1970's and 1980's~\cite{ANL_Barish_1977, BNL_Baker_1981, Kitagaki:1983px}
 with $O(10^3)$ events at most,
 on deuterium targets with nuclear corrections that are likely underestimated~\cite{Meyer:2016oeg}.
In addition, the dipole parameterization violates
 unitarity constraints imposed by QCD~\cite{Bhattacharya:2011ah},
 and although it has the expected asymptotic behavior at high-$Q^2$,
 this occurs for a regime well outside of the kinematic range probed
 by neutrino scattering experiments.

Figure~\ref{fig:gaq2_overlay} shows the status of existing calculations of
 the nucleon axial form factor from LQCD,
 compared to the axial form factor obtained from neutrino scattering
 with deuterium in Ref.~\cite{Meyer:2016oeg}.
The RQCD~\cite{RQCD:2019jai} and NME~\cite{Park:2021ypf} collaborations
 have the most mature analyses with several ensembles that probe a range of
 systematic effects.
These computations each have their own methods for addressing the
 excited state contamination discussed in Section~\ref{sec:lqcd_pcac},
 which successfully restore the GGT relation.
RQCD modify the parameterization used to fit the correlation function
 to better constrain the expected shape of excited state contamination from the $N\pi$ states based upon expectations from $\chi$PT~\cite{Bar:2018xyi}.
NME test a variety of Bayesian fits to constrain the excited state contributions,
 where their preferred fit enforces a tight prior on the nucleon-pion state
 at the energy expected from a naive dispersion relation.
Because these two results are based on several ensembles,
 their fits are parameterized and plotted as bands rather than as scatter points
 to distinguish them from estimates on single ensembles.

The ETMC~\cite{Alexandrou:2020okk}, LHPC~\cite{Hasan:2017wwt},
 PACS~\cite{Ishikawa:2018rew,Shintani:2018ozy,Ishikawa:2021eut}, and CalLat~\cite{Meyer:2021vfq}
 results have just a few ensembles, so scatter points obtained from fitting
 are shown rather than the form factor parameterization to distinguish
 them from extrapolated results.
Though these results are expected to be close to the physical point results,
 they will have unquantified systematic shifts.
The ETMC has three ensembles, two of which have only 2 flavors of sea quarks
 and will be subject to systematics from neglecting sea effects from strange quarks.
The form factor data from the remaining ETMC ensemble,
 a physical pion mass ensemble with 4 flavors of sea quarks (up, down, strange and charm),
 is shown in Fig.~\ref{fig:gaq2_overlay}.
The PACS results are on two different ensembles with physical pion mass,
 with volumes of $(5.5~{\rm fm})^3$ and $(10.8~{\rm fm})^3$.
The Mainz collaboration has an ongoing calculation in proceedings on 12 ensembles,
 including an ensemble at the physical pion mass
 and a chiral-continuum and infinite volume extrapolation~\cite{Djukanovic:2021yqg}.
The results from a two-state fit to their physical pion mass ensemble
 are plotted with the other results.%
%-------------------------------------------------------------------------------
\begin{marginnote}
\entry{Chiral-continuum}{A simultaneous extrapolation in the pion (quark) mass and the continuum limit}
\end{marginnote}%---------------------------------------------------------------
The CalLat data will be described in more detail in Sec.~\ref{sec:callatdata}.

The published results shown in Figure~\ref{fig:gaq2_overlay} use different lattice actions for the simulations ---
RQCD~\cite{RQCD:2019jai}, Mainz~\cite{Djukanovic:2021yqg} and PACS~\cite{Ishikawa:2018rew,Shintani:2018ozy,Ishikawa:2021eut} use non-perturbatively $\mathrm{O}(a)$ improved clover-Wilson fermions for the action and current (PACS notes that the improvement coefficient for the current is consistent with zero so they do not improve it),
NME~\cite{Park:2021ypf} and LHPC~\cite{Hasan:2017wwt} use tree-level (perturbative) $\mathrm{O}(a)$ improved clover-Wilson fermions for the action and no improvement for the current,
 ETMC~\cite{Alexandrou:2020okk} use twisted mass fermions (that are automatically $\mathrm{O}(a)$ improved) with a clover term,
 and CalLat~\cite{Meyer:2021vfq} uses a mixed-action setup with M\"obius DWF
 on HISQ~\cite{MILC:2012znn}%
 \begin{marginnote}
  \entry{HISQ}{Highly-improved staggered quarks}
 \end{marginnote}%
 gauge configurations (with leading scaling violations of $\mathrm{O}(a^2)$).
The general agreement between calculations with different actions tests
 the universality of fermion actions.
No obvious tensions are seen,
 leading to the conclusion that there are no significant scaling violations
 in the data due to nonzero lattice spacing.
The restriction to finite volume has also been probed to some extent by
 the PACS collaboration results,
 again with no obvious deviations from the results of other collaborations.

The excellent agreement of the axial form factor data and parameterizations
 for all of the LQCD simulations provides credibility to the claims
 made by the lattice collaborations, in particular the slow falloff with $Q^2$.
Despite the apparent violations of GGT in the low-$Q^2$ region,
 the high-$Q^2$ region seems to be in better control and not as sensitive to
 the same excited state contamination.
The high-$Q^2$ agreement is reflected by the restoration of the GGT relation at large $Q^2$,
 which is generally the case even for computations that have difficulty satisfying
 the GGT relation at low $Q^2$.
These claims could be modified by systematic effects
 that are common to all of the calculations,
 and excited state contaminations from $N\pi$ states in
 the axial form factor remain as a dominant concern.
While estimates of excited states using methods that are currently employed
 by lattice calculations have helped to clarify the situation,
 modern calculations with $N\pi$-like interpolating operators
 are needed to definitively quantify the excited state contaminations over all $Q^2$.
If a dedicated calculation can demonstrate that the
 excited states are controlled well by the methods presently in use,
 then worries about the contamination should be more or less resolved.

Another concern is that the magnitude of $Q^2$ may adversely affect
 the convergence of the chiral expansion at large $Q^2$,
 limiting the ability to extrapolate LQCD results to the physical point.
Even for low and moderate values of momentum transfer,
 a large expansion order would be needed to constrain the form factor
 dependence on the relevant low energy constants,
 limiting the predictability of the theory.
There is some hope that expanding in terms of the $z$ expansion parameter $z$
 may alleviate some of these concerns by building in correlations between
 low and high orders of $Q^2$ that are expected from analyticity.
This will be discussed in Section~\ref{sec:z_continuum},
 where the relationship between $Q^2$ and $z$ will be analyzed in more detail.

\subsubsection{Additional Results}
In addition to the aforementioned published results, there are
 a handful of recent preliminary results that deserve mention.
The LHPC~\cite{Hasan:2017wwt} and PACS~\cite{Ishikawa:2021eut} collaborations
 explore methods for directly computing the form factor values and slopes
 directly at $Q^2=0$, which offer alternative methods for constraining
 the form factor shape that could be used to complement traditional methods.
The Fermilab Lattice and MILC collaborations~\cite{Meyer:2016kwb,Lin:2019pia,Lin:2020wko} also have an ongoing
 computation of the axial form factor using a unitary HISQ-on-HISQ setup,
 for which a preliminary computation of the axial coupling on
 a single unphysical ensemble exists.
Because of the choice of action, this computation has more nucleon ``tastes''
 than other efforts, which is more computationally affordable
 at the cost of a more challenging analysis.

\subsubsection{Description of CalLat Data}\label{sec:callatdata}

Since the preliminary CalLat results~\cite{Meyer:2021vfq} for the axial form factor will be used
 in Section~\ref{sec:impact}, more discussion about these results is warranted.
The CalLat data are collected on a single ensemble generated
 by the MILC collaboration~\cite{MILC:2012znn}
 with a lattice spacing $a\approx 0.12~{\rm fm}$ and $m_\pi\approx 130~{\rm MeV}$.
Two-point correlation functions are computed with conjugate
 source and sink interpolating operators
 to produce a positive-definite correlation function.
The same source and sink operators are used for the three-point functions
 as well as a local insertion of the ${\cal A}_z$ axial current.
Up to 10 source-sink separation times were used in the range $t/a\in\{3,\dots,12\}$.

The setup used in this analysis fixes $\bm{q}$ at the current
 and projects the sink to 0 momentum, allowing the source momentum to be
 fixed by momentum conservation.
The momentum $\bm{q}$ at the insertion is chosen to have $q_z=0$,
 which explicitly zeros out all of the contribution to the correlator
 from the induced pseudoscalar.
Momenta up to $|q_{x,y}| \leq 4\sqrt{2}\cdot (2\pi/L)$ were explored,
 which corresponds to momentum transfers up to around $1~{\rm GeV}$.
The ground state axial matrix element is then proportional to the axial form factor,
 up to a known kinematic factor.
The correlator data are fit using a parameterization that allows
 for 3 states at each momentum.

Once the axial matrix elements were obtained,
 the form factor data were fit to a $5+4$-parameter $z$ expansion,
 including four parameters to enforce sum rules that regulate the large-$Q^2$ behavior,
\begin{align}
 \left( \frac{\partial}{\partial z} \right)^n
 \sum_{k=0}^{k_{\text{max}}+4} a_k z^k \Big|_{z=1} = 0; \; n \in \{0,1,2,3\}.
\end{align}
A prior was given to each of the $z$ expansion coefficients of the form
\begin{align}
 {\rm prior}\Big[ \frac{a_k}{|a_0|} \Big] = 0 \pm {\rm min} \Big[ 5, \frac{25}{k} \Big].
\end{align}
No attempt was made to correct for systematics due to lattice spacing,
 the pion mass mistuning, or the restriction to finite volume,
 and the uncertainties are statistical only.

The axial form factor coefficients obtained from this procedure are
 (omitting the sum rule coefficients)
\begin{align}
 a_k = \left\{ 0.914(10), -1.931(54), -0.63(30), 4.4(1.7), -2.2(3.6) \right\};
 \; k \in \{0,1,2,3,4\},
\end{align}
which are used in Section~\ref{sec:impact} without considering their uncertainties.

% ------------------------------------------------------------------------------
% z-expansion
\subsection{Combining the $z$ expansion with the continuum and chiral extrapolations\label{sec:z_continuum}}

Consider an expansion of the form factor as a power-series in $Q^2$ with coefficients that capture the pion mass, lattice spacing and lattice volume dependence
\begin{align}\label{eq:F_Q_power}
F(Q^2) = \sum_{k=0} f_k(m_\pi, a, L) Q^{2k}.
\end{align}
This expression will be valid for arbitrarily small values of $m_\pi$, $Q$ and $a$, and for sufficiently large values of $L$.
However, it is not valid for large values of $Q$: perturbative QCD predicts the form factor should scale as $Q^{-4}$ in the asymptotically large $Q^2$-regime.
For sufficiently small values of $m_\pi$ and $Q$ (and large $L$), $\chi$PT~\cite{Gasser:1984gg,Jenkins:1990jv,Bernard:1995dp} provides a model independent parameterization of the coefficients, $f_k$.  $\chi$PT can be extended to incorporate the lattice spacing corrections as well~\cite{Sharpe:1998xm}.

However, the LQCD calculations of the form factors are typically carried out up to $Q^2\approx1-3$~GeV$^2$, well outside the range of validity of $\chi$PT,%
% FOOTNOTE ---------------------------------------------------------------
\footnote{The range of validity of $\chi$PT in the pion mass seems to be at best $m_\pi\lesssim300$~MeV~\cite{Beane:2004ks,Walker-Loud:2008rui} with some indications the convergence is troubled at the physical pion mass~\cite{Walker-Loud:2019cif,Drischler:2019xuo}.  A similar upper limit in $Q$ is likely.}
%------------------------------------------------------------------------------
complicating a combined extrapolation in the various variables of interest.
For quantities such as the spectrum and nucleon axial, scalar and tensor couplings, results at heavier pion masses can help improve the overall precision of the final result provided the extrapolation to the physical pion mass is under control.
For the form factors,
 there is still an issue related to the continuum extrapolation even when using only ensembles with physical pion masses and large volumes with negligible volume corrections.
The reason is that the spatial extent, $L$, is given by the lattice spacing times
 the number of spatial sites, $N$.
In practice, given the non-trivial relation between the lattice spacing $a$, a derived quantity, and the bare gauge coupling that is an input parameter of the calculation, it is not possible to change $a$ in such a way that $L = N a$ is constant.
The allowed quantized momentum for the nucleon and currents will therefore change as the lattice spacing is changed, which means the value of the four-momentum transfer $Q$ will not be the same from one physical pion mass ensemble to the next.
Consequently, the continuum extrapolation, which is straightforward in principle, is complicated in practice, minimally requiring a combined extrapolation in $Q$ and $a$.
Is it possible to perform a combined extrapolation/interpolation analysis over the full range of individual $Q$ while also taking the continuum and physical pion mass limits?

The $z$ expansion takes advantage of the analytic structure of QCD
 by performing a conformal mapping to obtain a new small expansion parameter. This technique has been utilized for decades in meson flavor physics~\cite{Okubo:1971jf}
and is a standard feature in modern LQCD calculations of meson form factors
that constrain CKM matrix elements.
The $z$ expansion takes the four-momentum transfer squared $Q^2$
 to a small expansion parameter $z$, using the relation
\begin{align}
 z(t=-Q^2;t_0,t_c) = \frac{\sqrt{t_c-t} -\sqrt{t_c-t_0}}{ \sqrt{t_c-t} +\sqrt{t_c-t_0}}.
\end{align}
Both $t_0$ and $t_c$ are parameters that can be chosen, with some restrictions.
$t_c$ can not be larger than the particle production threshold in the timelike momentum transfer
 and $t_0$ is a parameter (typically negative) that
 may be chosen to improve the series convergence.
Inverting this relation and expanding as a power series in $z$ about $Q^2=-t_0$ ($z=0$) yields
\begin{align}
x \equiv  \frac{Q^2+t_0}{t_c-t_0} = 4 \sum_{k=1}^\infty k z^k.
 \label{eq:Q2toz}
\end{align}
Following this procedure, the dimensionful coefficients $f_k$ in equation~\eqref{eq:F_Q_power} are assembled into expressions related to the dimensionless coefficients of the $z$ expansion (which we denote here by $b_k$ to avoid confusion with the lattice spacing $a$)
\begin{align}
 F\big(z(Q^2; t_0, t_c)\big) = \sum_{k=0}^\infty b_k z^k.
 \label{eq:zexp}
\end{align}

The most recent multi-ensemble LQCD publications with computations
of the axial form factor~\cite{Park:2021ypf,RQCD:2019jai}
have treated the $z$ expansion coefficients as the relevant LECs
and fit to these coefficients with a chiral-continuum extrapolation.
As discussed above, for sufficiently small $m_\pi$, $Q$ and $a$, the coefficients $f_k$ have a well defined expansion that is rooted in $\chi$PT and its extension to incorporate discretization effects.%
%-------------------------------------------------------------------------------
\begin{marginnote}
\entry{LEC(s)}{Low energy constant(s), the coefficients of hadronic EFT operators}
\end{marginnote}%
%-------------------------------------------------------------------------------
This is possible because there is a rigorous mapping between the quark-level operators in QCD and the higher order Symanzik expansion with hadronic level operators in the EFT.
For a dynamic quantity, such as the form factor, or the $z$ expansion parameterization of it, it is not guaranteed that the coefficients can be described with simple $m_\pi$ and $a$ expansions.  For pion masses that are sufficiently close to the physical pion mass, and for lattice spacings sufficiently close to the continuum limit, a Taylor expansion about these limits should be adequate to interpolate and extrapolate with controlled uncertainties.
For some quantities, such as the proton radius, there are divergent $\ln(m_\pi)$ corrections that arise in the $\chi$PT description, which are observed as rapid pion mass dependence in LQCD calculations, see for example Ref.~\cite{Green:2014xba}.  Such corrections as these, or possibly important $a^2\ln(a)$ corrections that arise in lattice perturbation theory~\cite{Balog:2009yj,Balog:2009np,Husung:2019ytz}, if relevant, are ones that may not be easily or correctly captured by parameterizing the $z$ expansion coefficients as a power series in $a$ and/or $m_\pi$.
There is the additional subtlety of understanding and incorporating the finite volume corrections, which are well understood in the regime where $\chi$PT is applicable~\cite{Gasser:1986vb}, but we will not comment on in further detail.

We propose a strategy that might improve our systematic understanding of these effects, but needs to be explored to understand its validity and applicability.
In essence, the proposal is to begin with the EFT parameterization of the form factor and perform the conformal mapping to express the $z$ expansion coefficients, $b_k$ in terms the $f_k$ coefficients that are derived from $\chi$PT and its finite volume and lattice spacing EFT extensions.
Then, one can compare the values of the LECs determined in the $z$ expansion analysis over the entire range of data, to an analysis of the form factor in the low $m_\pi$ and low $Q^2$ region utilizing the $\chi$PT expression.  One could also explore simultaneous fits with both methods, where one utilizes higher order terms in the $z$ expansion, if needed, which could also help quantify corrections that cannot be systematically described with the $\chi$PT parameterization.

As a slightly more concrete example, start with the inverse transformation from $z$ back to $Q^2$, expressed in terms of $x$, given in Equation~\eqref{eq:Q2toz}.  The expression for $z$ at small $x$ is \begin{align}
 (1+x)^{1/2} -1 = \frac{x}{2}
 -4\sum_{k=2}^\infty \frac{(2k-3)!}{k!(k-2)!} \biggr( -\frac{x}{4} \biggr)^{k},
 &\quad
 z = \frac{1}{x} \big( (1+x)^{1/2} -1 \big)^2,
 \label{eq:ztoQ2}
\end{align}
 which starts at $O(x)$.
A general series in $Q^2$ may therefore be converted to a double expansion in $z$ and $t_c-t_0$
 by first converting powers of $Q^2$ to those of $Q^2+t_0$ and $t_0$ using the  binomial theorem,
\begin{align}
 Q^{2m} &= \big( (Q^2+t_0) -t_0 \big)^m
 = (t_c-t_0)^m
 \sum_{n=0}^{m} \left( \begin{array}{c} m \\ n \end{array} \right)
 x^n
 \biggr( \frac{-t_0}{t_c-t_0} \biggr)^{m-n}
\end{align}
and then substituting Equation~(\ref{eq:Q2toz}) to convert powers of
 $x$ into powers of $z$.
All dependence on the dimension is absorbed into powers of
 $t_c-t_0 \sim m_\pi^2$.
The relative weight of the expansion parameters may be adjusted by changing the value of $t_0$, giving some modicum of freedom over the expansion order.
If we consider the $Q^4$ expansion of Equation~\eqref{eq:F_Q_power} with
 a truncation at the leading chiral and discretization corrections, we get
\begin{align}
f_0 &= c_0 + \ell_0 m_\pi^2 + d_0 a^2\, ,
\nonumber\\
f_1 &= c_1 + \ell_1 m_\pi^2 + d_1 a^2\, ,
\nonumber\\
f_2 &= c_2 + \ell_2 m_\pi^2 + d_2 a^2\, ,
\label{eq:lecampi}
\end{align}
 where $c_k$, $\ell_k$ and $d_k$ are $\chi$PT expressions, including non-analytic $\ln(m_\pi)$ type and LEC corrections describing the pion mass and lattice spacing dependence.
For the axial form factor, $c_0$ and $c_1$ are related to the axial coupling and radius in the chiral limit
\begin{align}
&c_0 = \lim_{m_\pi\rightarrow0} g_{\mathrm{A}}\, ,&
&c_1 = -\lim_{m_\pi\rightarrow0} \frac{g_{\mathrm{A}} r_{\mathrm{A}}^2}{6}\, .&
\end{align}
Then the $z$ expansion coefficients that appear in Equation~(\ref{eq:zexp}),
 expressed in terms of $f_k$, are
\begin{align}\label{eq:z_coeff_xpt}
b_0 &= f_0 +\tctza \tctzb f_1 +\tctza^2 \tctzb^2 f_2 +O\big(\tctza^3\big),
\nonumber\\
b_1 &= 4 \tctza f_1 +8 \tctza^2 \tctzb f_2 +O\big(\tctza^3\big),
\nonumber\\
b_2 &= 8 \tctza f_1 +16 \tctza^2 \Big(1 -\frac{t_0}{t_c-t_0}\Big) f_2 +O\big(\tctza^3\big).
\end{align}
The leading lattice spacing, pion mass and finite volume dependence may be made manifest by substituting their expressions from Equation~(\ref{eq:lecampi}).

% ------------------------------------------------------------------------------
% Impact
\section{Phenomenological Impact\label{sec:impact}}

Neutrino oscillation experiments measure an event rate, which is the convolution of the flux, cross section and detector efficiency, as a function of some measurable variable. The incoming neutrino energy is not known event by event, and not all outgoing particles are detectable, so quantities such as the energy transfer, or four-momentum transfer, cannot be reconstructed. As neutrino oscillation is a neutrino energy (and distance) dependent phenomenon, experiments attempt to reconstruct it using the kinematics of particles produced when neutrinos interact in their detectors.

T2K, and other experiments with a relatively low energy ($\lessapprox1$ GeV) beam~\cite{Hyper-Kamiokande:2018ofw, MiniBooNE:2020pnu}, attempt to reconstruct the neutrino energy using outgoing lepton momentum, $p_{l}$, and its angle with respect to the incoming beam direction, $\theta_{l}$, assuming two-body quasielastic kinematics with the initial nucleon at rest,
\begin{equation}
E^{\mathrm{rec,\;QE}}_{\nu}\left(p_{l}, \theta_{l}\right) = \frac{2m_f\sqrt{p_{l}^2 + m^2_l} - m_l^2 + m_i^2-m_f^{2}}{2\left(m_f-\sqrt{p_{l}^2 + m^2_l}+p_l \cos\theta_l\right)},
\label{eq:enuqe}
\end{equation}
\noindent where $m_l$ is the mass of the outgoing lepton, $m_{i}$ is the mass of the initial state nucleon, and $m_{f}$ is the mass of the final state nucleon. As this variable assumes quasielastic kinematics, it is applied to a signal sample of CC0$\pi$ events.\footnote{Note that in recent analyses, T2K has included samples with a single charged pion using a modified version of Equation~\eqref{eq:enuqe}~\cite{T2K:2017rgv, T2K:2019bcf, T2K:2021xwb}.}%
\begin{marginnote}
\entry{CC0$\pi$}{Events with a muon, no pions or other mesons, and any number of nucleons produced in the final state}
\entry{CCQE}{Charged-current quasielastic}
\entry{CC-2p2h}{Charged-current interactions with two nucleons}
\entry{CC-RES}{Charged-current resonant pion production}
\end{marginnote}%
Events that are not true CCQE events also contribute to the CC0$\pi$ signal, such as CC-2p2h or CC-RES with no visible final-state pion. The two-body approximation in Equation~\eqref{eq:enuqe} is a poor approximation of the true neutrino energy, $E_{\nu}^{\mathrm{true}}$, in these cases. Understanding the relative fraction of the different interaction channels is therefore a critical issue for experiments that use Equation~\eqref{eq:enuqe}.

\begin{figure}[htbp]
  \centering
  \captionsetup[subfloat]{captionskip=-5pt}
  \subfloat[Near detector]{\includegraphics[width=0.3\textwidth]{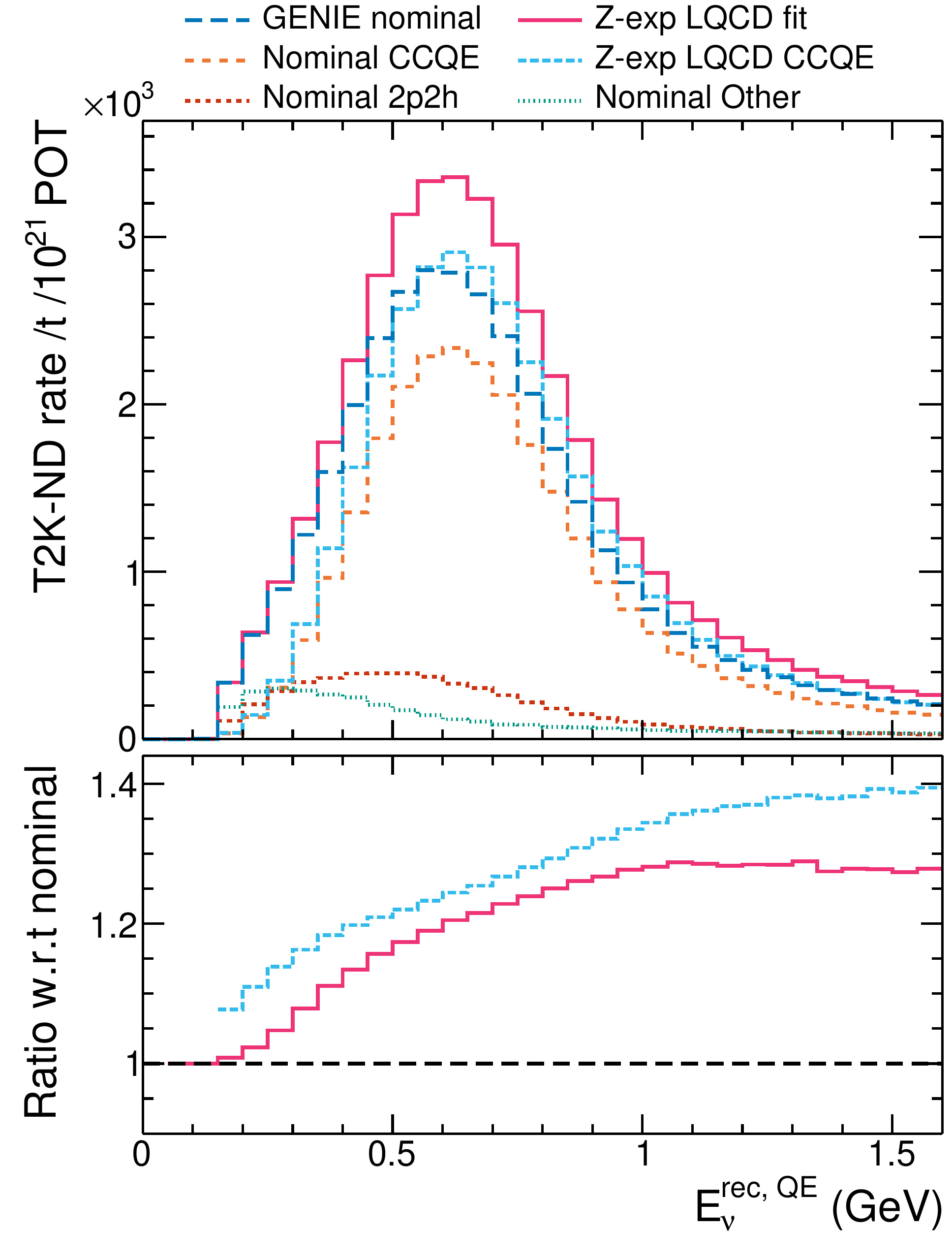}}\hspace{75pt}
  \subfloat[Far detector] {\includegraphics[width=0.3\textwidth]{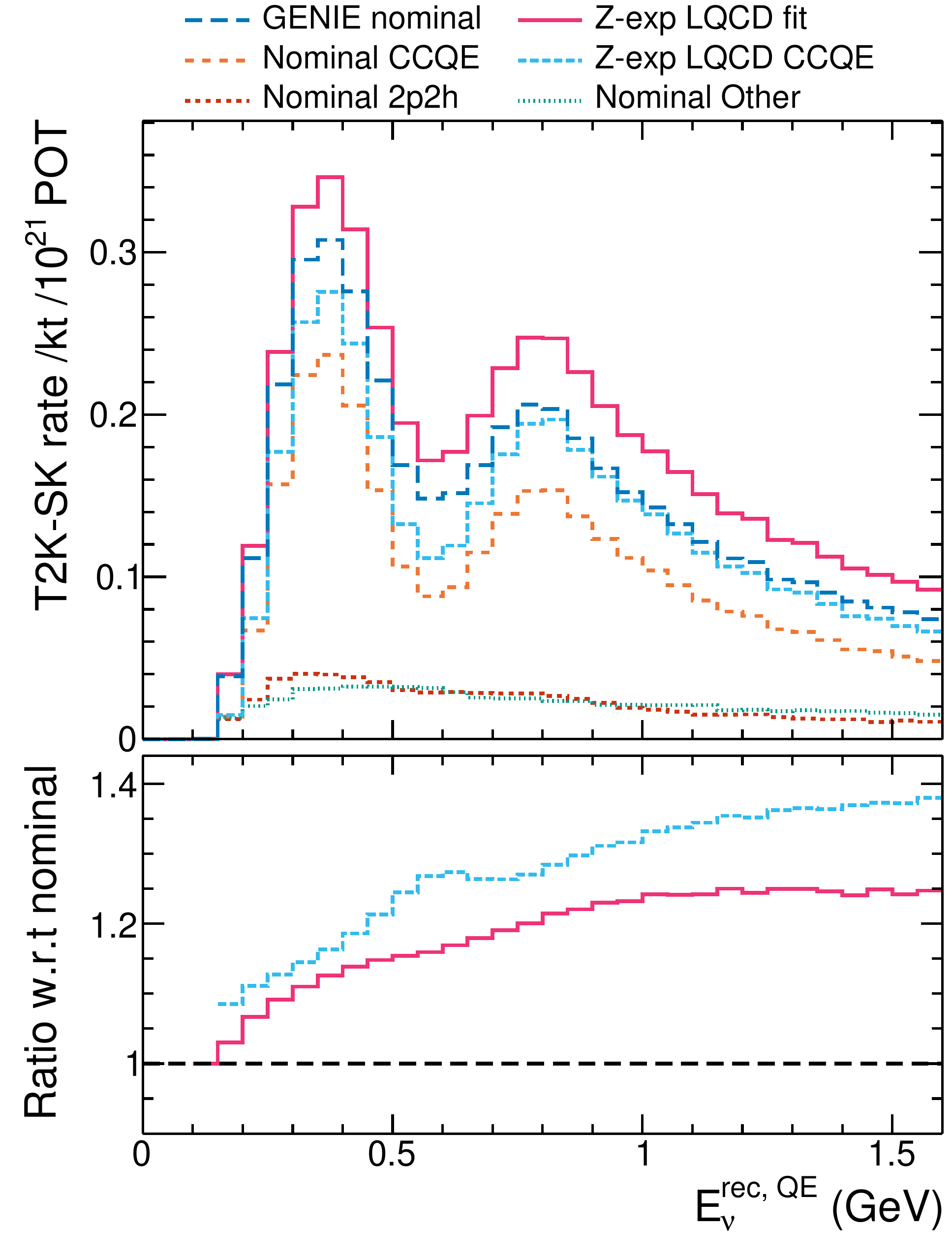}}
  \vspace{11pt}
  \caption{The $\nu_{\mu}$--H$_{2}$O CC0$\pi$ event rates per ton (kiloton) per $1\times10^{21}$POT at T2K's near (far) detector site, shown as a function of $E^{\mathrm{rec,\;QE}}_{\nu}$. The GENIE~\cite{Andreopoulos:2009rq, GENIE:2021npt} nominal event rate (blue solid line) is produced using the GENIEv3 10a\_02\_11a tune to nucleon data~\cite{GENIE:2021zuu} and the T2K flux~\cite{T2K:2012bge}. The CCQE (orange dashed line), CC-2p2h (red short dashed line) and CC-other (green dotted line), here meaning all events that are not CCQE or CC-2p2h, contributions are shown. The oscillated flux is calculated using the best fit NuFit5.0 oscillation parameters in normal ordering~\cite{Esteban:2020cvm, nufitweb}. Additionally, an alternative GENIE model is shown, where the only change is to use the $z$ expansion model of the axial form factor, with parameters tuned to LQCD results from the CalLat collaboration, as described in Section~\ref{sec:callatdata}. The ratio of the modified to nominal GENIE models is shown in the bottom panel of each plot.}
  \label{fig:t2k_impact}
\end{figure}
Figure~\ref{fig:t2k_impact} shows the $\nu_{\mu}$--H$_{2}$O CC0$\pi$ event rate expected at the T2K near and far detectors for a fixed exposure, shown as a function of $E^{\mathrm{rec,\;QE}}_{\nu}\left(p_{l}, \theta_{l}\right)$, with and without modifications to the axial form factor. The nominal GENIEv3 10a\_02\_11a model~\cite{Andreopoulos:2009rq, GENIE:2021npt} uses a dipole axial form factor with $M_{\mathrm{A}} = 0.941$%
\begin{marginnote}
 \entry{$M_{\mathrm{A}}$}{dipole axial mass parameter}
\end{marginnote}%
 GeV obtained through a fit to bubble chamber data~\cite{GENIE:2021zuu}. The alternative model shown differs only in the use of the $z$ expansion model for the axial form factor, with parameters tuned to the LQCD results from the CalLat collaboration described in Section~\ref{sec:callatdata}.
One obvious observation to be made from Figure~\ref{fig:t2k_impact} is that the CCQE events
, for which the axial form factor is relevant,
make up the majority of T2K's CC0$\pi$ sample
in both the oscillated and unoscillated fluxes.
The tuned LQCD values have a significant effect on the total expected event rate in both cases, on the order of 20\%.

Also clear from Figure~\ref{fig:t2k_impact} is that the change in both the total and CCQE-contributed event rates is not purely a normalization change, but instead depends on neutrino energy and is different for the near and far detector.
In the T2K oscillation analysis, the near detector data is used to constrain the cross section model, to reduce the uncertainty at the far detector and on the measured oscillation parameters. If there is insufficient freedom in the cross section model to account for this change to the CCQE model, other parts of the model may well be distorted to ensure good agreement with the near detector data.
However, this can introduce biases at the far detector, as the different interaction types that contribute to the CC0$\pi$ samples shown in Figure~\ref{fig:t2k_impact} have very different $E_{\nu}^{\mathrm{true}}$ --- $E^{\mathrm{rec,\;QE}}_{\nu}$ relationships. So if other components of the model are distorted to fit the unoscillated near detector event rate, that same {\it effective} change is not in general expected to extrapolate to the far detector spectrum correctly.

Experiments with higher neutrino beam energies typically do not limit their analyses to CC0$\pi$ events or use Equation~\ref{eq:enuqe} for reconstructing the neutrino energy. Instead, they use all charged-current events (CC-inclusive), and reconstruct the neutrino energy through a combination of particle identification and tracking together with calorimetry.
For an ideal detector, with no tracking threshold on protons, charged pions or electromagnetic activity, this can be expressed as,
\begin{equation}
E_{\nu}^{\mathrm{rec,\;had}} = E_{l} + \Sigma_{p} E_{\mathrm{kin}} + \Sigma_{\pi^{\pm}, \pi^{0}, \gamma} E_{\mathrm{total}},
\label{eq:enuhad}
\end{equation}
\noindent where $E_{l}$ is the energy of the outgoing charged lepton, and $E_{\mathrm{kin}}$ and $E_{\mathrm{total}}$ indicate the kinetic and total energies of individual outgoing hadrons.
The $E_{\nu}^{\mathrm{true}}$ --- $E_{\nu}^{\mathrm{rec,\;had}}$ smearing in this idealized case is due to missing kinetic energy lost to neutrons, and initial state nuclear effects (e.g., nucleons are not at rest inside the nucleus). Real detectors have tracking thresholds below which charged particles cannot be reconstructed, which results in additional smearing due to missing the masses of charged pions, although some energy lost to neutrons may be recovered.

\begin{figure}[htbp]
  \centering
  \subfloat[ND]{\includegraphics[width=0.3\textwidth]{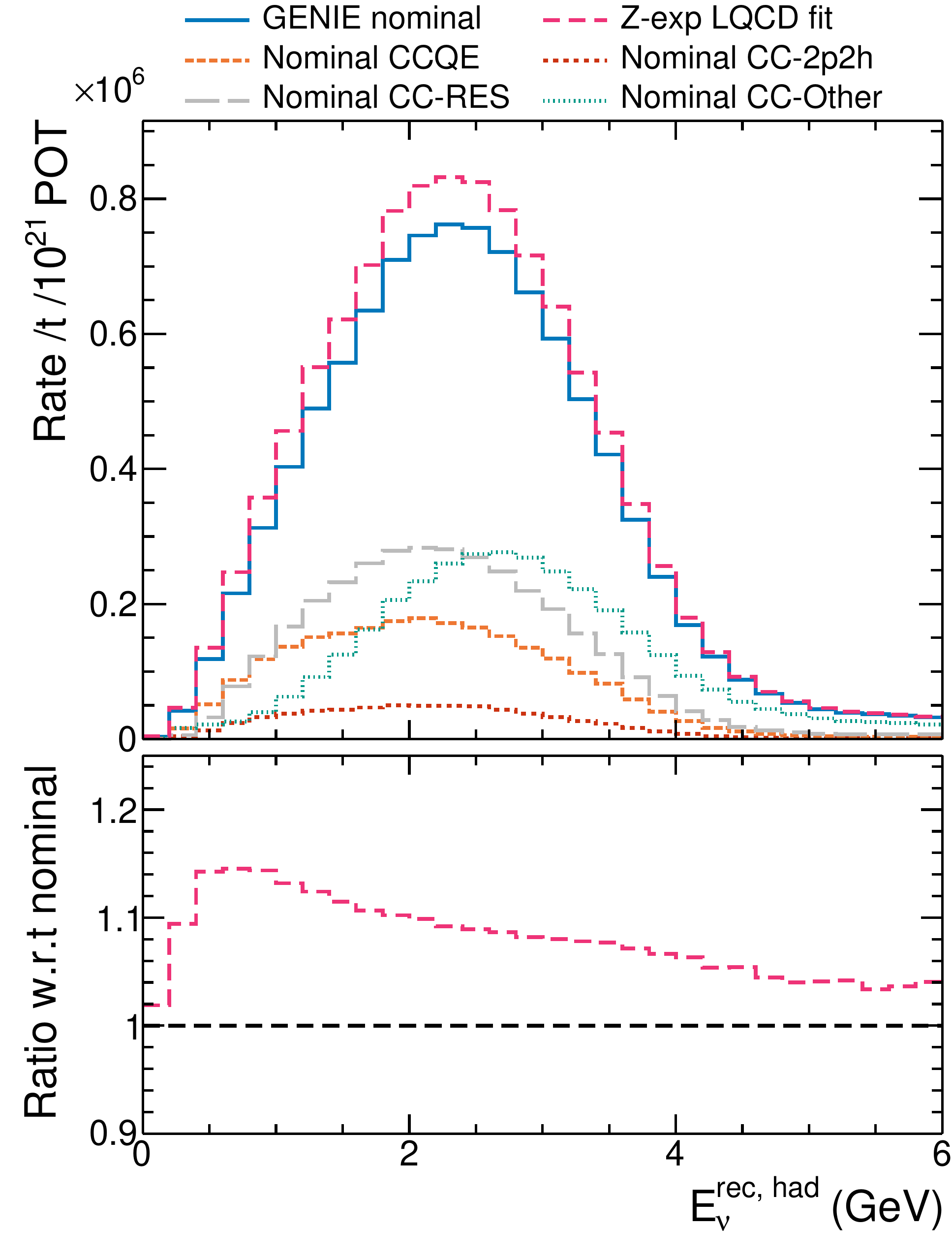}}\hspace{75pt}
  \subfloat[FD]{\includegraphics[width=0.3\textwidth]{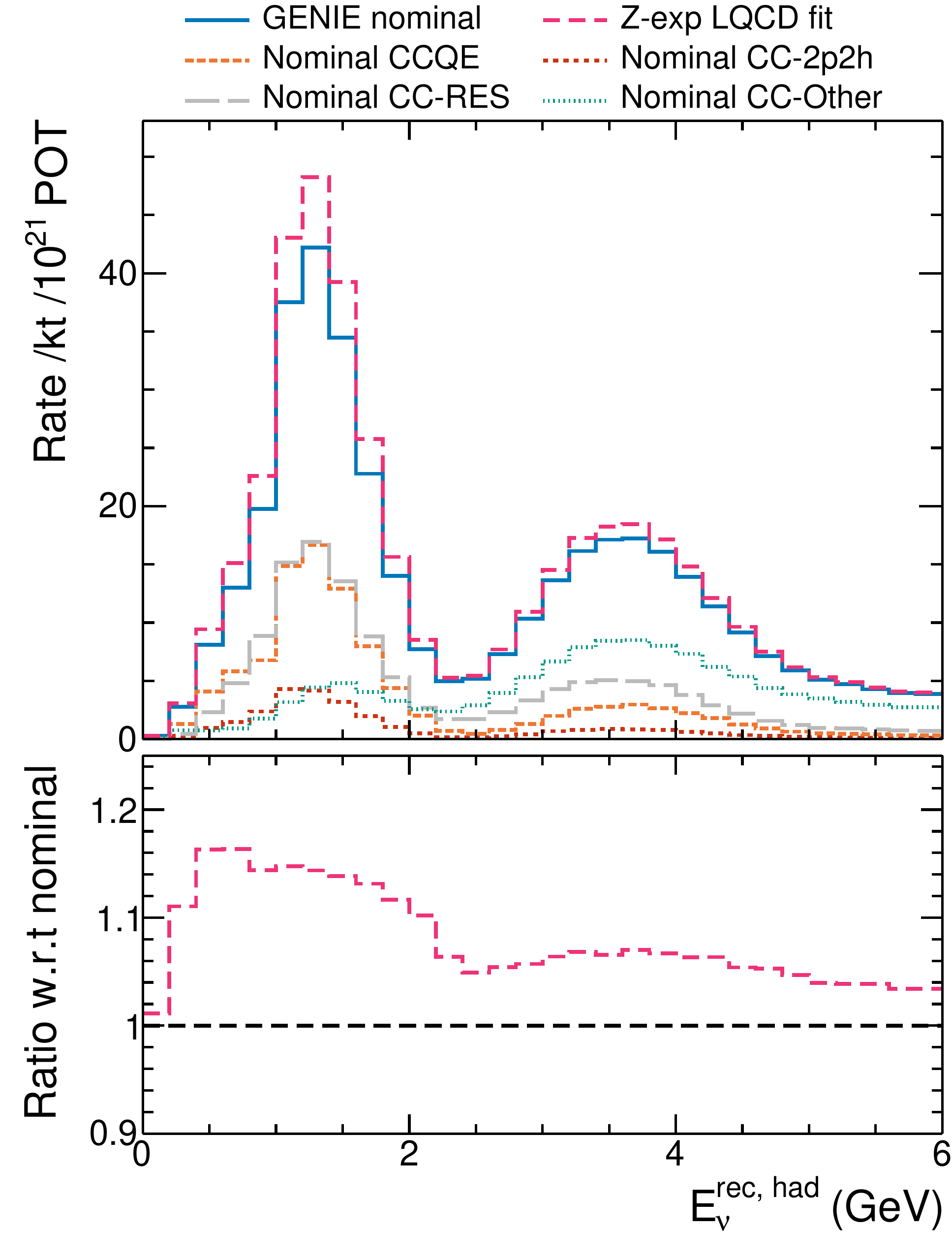}}
  \vspace{11pt}
  \caption{The $\nu_{\mu}$--$^{40}$Ar CC-inclusive event rates per ton (kiloton) per $1\times10^{21}$POT at DUNE's near (far) detector site, shown as a function of $E^{\mathrm{rec,\;had}}_{\nu}$. The GENIE~\cite{Andreopoulos:2009rq, GENIE:2021npt} nominal event rate (blue solid line) is produced using the GENIEv3 10a\_02\_11a tune to nucleon data~\cite{GENIE:2021zuu} and the DUNE flux~\cite{Abi:2020evt}. The CCQE (orange dashed line), CC-2p2h (red short dashed line), CC-RES (long-dashed gray line) and CC-other (green dotted line), here meaning all events that are not CCQE, CC-2p2h or CC-RES, contributions are shown. The oscillated flux is calculated using the best fit NuFit5.0 oscillation parameters in normal ordering~\cite{Esteban:2020cvm, nufitweb}. Additionally, an alternative GENIE model is shown, where the only change is to use the $z$ expansion model of the axial form factor, with parameters tuned to LQCD results from the CalLat collaboration, as described in Section~\ref{sec:callatdata}.
The ratio of the modified to nominal GENIE models is shown in the bottom panel of each plot.}
  \label{fig:dune_impact}
\end{figure}
Figure~\ref{fig:dune_impact} shows the $\nu_{\mu}$--$^{40}$Ar CC-inclusive event rates per ton (kiloton) per $1\times10^{21}$POT at DUNE's near (far) detector site, shown as a function of $E^{\mathrm{rec,\;had}}_{\nu}$, with and without modifications to the axial form factor. At this higher neutrino beam energy (with $E_{\nu}^{\mathrm{peak}} \approxeq 2.5$ GeV), CCQE events still make up a sizable, $\approx$30\%, fraction of the total events.
The modification to the axial form factor based on the LQCD results from the CalLat collaboration described in Section~\ref{sec:callatdata} results in an approximately 10\% enhancement to the total predicted event rate at both the near and far detectors. As in the case with the T2K event rate, the enhancement has a nontrivial neutrino energy dependence.
Despite the different neutrino energy reconstruction methods used by DUNE and T2K, the same arguments about potential bias due to model-dependence apply when there are differences in the effect of an out-of-model change between near and far detectors.

The issue of how to assign strength between the CCQE and CC-2p2h channels has been a major focus for neutrino oscillation experiments over the past decade.
Data from a large number of experiments has disagreed with model predictions in the CC$0\pi$ channel at the 10--30\% level~\cite{garvey_review_2014, Mosel:2016cwa, NuSTEC:2017hzk, Katori:2016yel, ParticleDataGroup:2020ssz}. This has prompted development of {\it ad hoc} systematic uncertainties and empirical model tunings unique to each experiment, with a tendency to soak up model-data discrepancies into the CC-2p2h channel.
These are major contributors to the final uncertainties on key oscillation parameter measurements and projected sensitivities~\cite{T2K:2019bcf, DUNE:2020jqi, T2K:2021xwb, NOvA:2021nfi, DUNE:2021mtg}. It is therefore very significant that the LQCD results shown in Figures~\ref{fig:t2k_impact} and~\ref{fig:dune_impact} suggest that an increase to the strength of the CCQE contribution on the order of 20\% is necessary.

% ------------------------------------------------------------------------------
% Future
\section{Future Improvements\label{sec:future}}

The most pressing issue that needs to be definitively resolved for the LQCD calculations is whether or not the excited state contamination is under complete control for the nucleon (quasi-)elastic form factor calculations.
If they are, then the LQCD results imply that the nucleon axial form factor is significantly different than what has been extracted phenomenologically, as indicated in Figure~\ref{fig:gaq2_overlay}.
However, it is worth observing that LQCD calculations of $g_{\mathrm{A}}$ were systematically low for many years, before it was finally understood the issues was related to an under-estimation of the systematic uncertainty associated with these excited states.

There are several groups computing these form factors, with several different lattice actions and several different approaches to quantifying and removing the excited state contamination from the ground state matrix elements.
Given the heightened awareness of this issue, it is much less likely that all the LQCD results are polluted by such a contamination than was the case for $g_{\mathrm{A}}$.
The most clear way to definitively resolve the question would be to perform an extremely high statistics calculation with source-sink separation times of $\tsep\approx$2--3~fm.  The extreme numerical cost
renders this an unlikely approach.
In the longer term, the use of multi-level integration schemes can lead to an exponentially improved stochastic precision~\cite{Ce:2016idq}.

A much more practical solution presently would be to implement a variational calculation using distillation~\cite{HadronSpectrum:2009krc} or its stochastic variant~\cite{Morningstar:2011ka}.
There are several advantages to such a calculation.
First, these methods enable the use of multi-hadron creation and annihilation operators
 that are essential to properly identify both the spectrum and the nature of the state~\cite{Dudek:2012xn,Lang:2012db},
 whether it is a $P$-wave $N\pi$ state,
 some radial excitation of the nucleon such as the Roper
 that prominently decays to an $N\pi\pi$ state,
 or otherwise.
Given this information, one can construct linear combinations of these operators which systematically remove the excited states from the correlation function~\cite{Blossier:2009kd}, allowing for the utilization of much earlier Euclidean times where the stochastic noise is relatively much smaller.
Second, these variational methods enable the use of momentum space creation as well as annihilation operators, with full control over the spin projection of the source and sink at minimal extra cost.
This enables one to construct useful linear combinations of correlation functions that eliminate, for example, the induced pseudoscalar contribution to an axial three point function, which simplifies the analysis~\cite{Meyer:2017ddy}.
Further, one can exploit the Breit-Frame in which magnitude of the incoming and outgoing momentum are the same which opens the door for utilizing the Feynman-Hellmann-like correlation functions which suppress the excited state contamination significantly compared to the three-point functions~\cite{He:2021yvm}.
It is encouraging that the first exploratory calculations with such methods have begun~\cite{Egerer:2018xgu,Barca:2021iak}.

Moving beyond these simplest quantities, higher energy transfer processes also play a sub-dominant role for T2K (and the future
Hyper-K) experiments, and a major role in the DUNE experiment which has a higher energy neutrino
flux, as can be seen in Figures~\ref{fig:t2k_impact} and~\ref{fig:dune_impact}, respectively.
These higher energy transfers can access other fundamentally different interaction topologies,
 such as resonant or nonresonant pion production mechanisms,
 nuclear responses with correlated nucleon pairs,
 or scattering off partons within nucleons.
In principle, all of these interaction mechanisms are accessible to LQCD,
 though with varying degrees of difficulty.
Given the discrepancy between lattice axial form factor data and experimental constraints,
 it is not unreasonable to expect other interaction mechanisms have similar discrepancies
 between theory and observation.
These interaction types are as inaccessible as CCQE with modern $\nu A$ data, typically relying on old H$_2$ and D$_2$ bubble chamber datasets, with even lower statistics than the historic CCQE datasets already discussed, compounding the problem.
Appeals to model assumptions may give some handle for missing quantities,
 but are also subject to unquantifiable systematic effects.

Calculations that access the combined resonant and nonresonant scattering amplitudes
 are the most similar to those of elastic scattering,
 where a current induces a transition of the nucleon to a multiparticle final state.
The most prominent resonant contribution is from the $N\rightarrow\D$ transition.
However, building up an understanding of the entire resonant region, following the standard LQCD computational strategy, will be an extremely challenging endeavor given the dense spectrum of multi-particle states, and more importantly, the lack of formalism to relate three-particle matrix elements in finite volume to the infinite volume physics of interest.
Alternative strategies that focus on the inclusive $N\rightarrow X$ contribution~\cite{Hansen:2017mnd,Gambino:2020crt,Fukaya:2020wpp,Bruno:2020kyl}, or the use of two-currents to compute the hadronic tensor~\cite{Liu:1993cv,Liang:2019frk} are likely to be more fruitful.

Calculations of two-nucleon matrix elements provide key insights
 about the correlations between nucleons inside a nuclear medium,
 a vital ingredient for construction of an effective theory
 of $\nu A$ interactions.
The first efforts have been made to compute two-nucleon matrix elements in response to electroweak currents~\cite{Savage:2016kon,Chang:2017eiq}.
These calculations will most likely have to be revisited with a variational method as well,
 since it now appears the use of local two-nucleon creation operators do not correctly
 reproduce the spectrum~\cite{Francis:2018qch,Horz:2020zvv,Green:2021qol,Amarasinghe:2021lqa}.
Future calculations of matrix elements for currents inserted between
 two-nucleon states could provide direct information about the LEC inputs to EFT descriptions of nuclear physics.
While EFT can not be used to describe the $\nu A$ response over the full kinematic range of interest, they can provide a crucial anchor point to constrain nuclear models, see for example Refs.~\cite{Kronfeld:2019nfb,Drischler:2019xuo,Tews:2020hgp,Davoudi:2020ngi}.
As a specific example, deuterium corrections from nuclear models were assumed to be strong only at low momentum transfer
 and energy-independent in the reanalysis of deuterium bubble chamber data~\cite{Meyer:2016oeg},
 despite the inability of these corrections to account for the theory-data discrepancies.
A direct LQCD computation of these effects would isolate the effect,
 either by definitively attributing the discrepancy to deuterium effects
 or by implicating the other systematics corrections.

% ------------------------------------------------------------------------------
% Conclusions
\section{Conclusions\label{sec:conclusions}}

LQCD collaborations are able to produce consistent results for benchmark quantities such as $g_{\mathrm{A}}$ with percent level systematic uncertainties, which are in excellent agreement with experimental data.
These results introduce the exciting possibility of using LQCD calculations to tackle other quantities that are not easily accessible experimentally, or for which tensions between measurements, or competing models, exist.
In this review, we discussed LQCD calculations of nucleon form factors as a function of momentum transfer, which are of particular interest to the few-GeV neutrino experimental program.
Here we focused on the axial form factor, $F_{\mathrm{A}}(Q^2)$, which is of primary importance because current parameterizations are simplistic and rely on a handful of low-statistics $\nu N$ bubble chamber measurements.
However, notable tensions exist in current parameterizations of the vector form factors too.
$F_{\mathrm{A}}(Q^2)$ cannot be cleanly measured with existing experiments that use heavier nuclear targets both for safety reasons and to increase the event rate, so LQCD offers a novel path to this important quantity.
We have compared $F_{\mathrm{A}}(Q^2)$ calculations from a variety of different LQCD collaborations using different approaches and techniques, and shown them to be in good agreement with each other, but crucially, in poor agreement with the simple dipole model tuned to historic $\nu N$ data that are currently relied upon.
Assuming that no systematic effects affecting all of the LQCD calculations are uncovered, this suggests a significant increase of approximately 20\% to the strength of the CCQE scattering channel that dominates the neutrino scattering cross section for $E_{\nu} \lesssim 1$ GeV.

While it is possible that the lattice community will uncover additional systematic uncertainties in the calculation of the axial form factor, given the current state of understanding in the field, we believe this is unlikely.  The consistency of the various incomplete LQCD results at the physical pion mass lend confidence to this assessment, and these results will continue to be refined and substantiated.
Given the state of the field and the rapid advances recently made, we anticipate that the systematic uncertainties will be fully understood within a year or two,
 thus enabling precise determinations of the quasielastic axial form factor and the elastic electric and magnetic form factors.
At this point, LQCD will be able to resolve the experimental and phenomenological discrepancies reviewed in Section~\ref{sec:sof}.

We have demonstrated that these results produce a significant change in the predicted neutrino event spectra for T2K (which has similar considerations to Hyper-K) and DUNE experiments.
Determining the impact on oscillation results would require a full analysis performed by each experimental collaboration, but it is clear that LQCD results for $F_{\mathrm{A}}(Q^2)$ may offer a valuable insight that can clarify aspects of the complex neutrino interaction modeling problem these experiments face.
We additionally discussed a number of ways that current calculations can be improved and validated, and to increase confidence that the LQCD results are not subject to an uncontrolled systematic uncertainty.
Finally, we discussed a number of other quantities that are important to neutrino oscillation experiments in the few-GeV energy regime and that LQCD can provide first principles predictions for.
These include resonant pion production at higher energy transfers (of particular interest to DUNE), and insights into nucleon-nucleon correlations, for which there are no clear experimental prospects.
These possibilities would all work to break current degeneracies between $\nu N$ interactions and various nuclear effects that all come into play for present modeling efforts of the $\nu A$ cross sections.

\bigskip\noindent
\textbf{Note added in proof:}
After this review was completed, the Mainz Collaboration updated their results presented in the Lattice Proceedings~\cite{Djukanovic:2021yqg} with an arXiv preprint containing a full extrapolation to the physical point, see Figure 3 or Ref.~\cite{Djukanovic:2022wru}.

%Disclosure
\section*{DISCLOSURE STATEMENT}
While the authors collaboration affiliations do not affect the objectivity of this review, we wish to include them for transparency. ASM and AWL are both current members of the CalLat collaboration. ASM is a current member of the Fermilab Lattice and MILC collaborations. CW is a current member of the T2K and DUNE collaborations.
The authors are not aware of any funding, or financial holdings that might be perceived as affecting the objectivity of this review.

% Acknowledgments
\section*{ACKNOWLEDGMENTS}
The work of ASM was supported by the Department of Energy, Office of Nuclear Physics, under Contract No. DE-SC00046548.
The work of AWL and CW was supported by the Director, Office of Science, Office of Basic Energy Sciences, of the U.S. Department of Energy under Contract No. DE-AC02-05CH11231.

% References
\bibliography{AR_review.bib}
\bibliographystyle{ar-style5}
%ArXiv references may be formatted as follows, in the Literature Cited section: “1. Author A, Author B. arXiv:XXXX.XXXX [hep-ph] (2017)”

\end{document}